\documentclass[showpacs,twocolumn,amssymb,aps,flushbottom,floatfix,balancelastpage]{revtex4-1}
\usepackage{dcolumn}
\usepackage{bm}
\usepackage{graphicx}
\usepackage{epstopdf}
\usepackage{amsmath}
\usepackage[dvipsnames,usenames]{color}
\usepackage{subfig}
\usepackage{color}
\usepackage[colorlinks=true]{hyperref}
\usepackage{cleveref}
\usepackage{natbib}
\newcommand{\RN}[1]{%
  \textup{\uppercase\expandafter{\romannumeral#1}}%
}
\usepackage[font=small,
   justification=justified,
   format=plain]{caption}
\begin{document}
\title{Higher-order modulation instability in a fourth-order Nonlinear Schr\"{o}dinger Equation}
\author{Amdad Chowdury}
\affiliation{School of Physical and Mathematical Sciences, Nanyang Technological University, Singapore 639798}
\begin{abstract}
We present a complete dynamical descriptions of the higher-order modulation instability for a fourth-order nonlinear Schr\"{o}dinger equation. For two-breather solutions of this equation, we have identified the locus in a geometrical space where the growth rates for the breathers are equal in a parameter space. We show that, for the two-breather solutions of nonlinear Schr\"{o}dinger equation, the entire parameter space is bounded by a circle, whereas it is bound by an intersecting circle and an ellipse for the fourth-order equation. We show that, for all the higher-order equation in the nonlinear Schr\"{o}dinger equation hierarchy, the parameter space follows similar geometric interpretation as for the fourth-order equation.
\end{abstract}
\pacs{05.45.Yv, 42.65.Tg, 42.81.Qb, 02.30.Ik}
\maketitle

\section{Introduction}

The phenomenon of modulation instability (MI) is  central to the dynamics of nonlinear evolution processes. In the initial stage of the development of MI, a periodic perturbation will amplify exponentially and generate cascades of spectral sidebands \cite{benjamin1967disintegration, bespalov1966filamentary}. In the developed stage of MI, the dynamics are more complex and involve several stages of energy exchange among the spectral modes. The evolutionary processes of MI can be explained using another similar phenomenon called Fermi-Pasta-Ulam (FPU) recurrence \cite{van2001experimental}. 
 
 In   nonlinear optics, MI has been the subject of extensive study because of its inherent connection with short pulse dynamics in nonlinear optical media \cite{tai1986observation}, such as generating a  train of ultrashort   pulses, and also optical parametric amplification \cite{greer1989generation, agrawal1987modulation} . Additionally, it was shown that the underlying mechanism involved in  the initial stage of supercontinuum generation development is connected to   noise-driven MI, which is usually seeded by  continuous wave radiation \cite{demircan2005supercontinuum,dudley2006supercontinuum,travers2008visible,dudley2009modulation,genty2010akhmediev}. 
 
Recently, it was also shown that the emergence of a rogue wave in the developed stage of supercontinuum generation (SCG) is also connected with MI \cite{Solli:07:Nature}. Mostly, the study of MI development in nonlinear optics was conducted using the basic nonlinear Schr\"{o}dinger equation (NLSE). Though the analytic solution of the NLSE called the  Akhmediev breather (AB) is  existing for more than 30 years \cite{akhmediev1986modulation}, and has the the ability to explains MI phenomena in nonlinear media, researchers in  similar fields mostly used numerical approaches to investigate MI \cite{trillo1991dynamics, trillo1994nonlinear}. Recently, the analytic solution of an AB \cite{akhmediev1986modulation} is being used for the analysis of the evolution of how a modulated continuous wave field transform into a train of ultrashort pulses \cite{dudley2009modulation,genty2010akhmediev}. In the same work, it was also shown how the AB provides insight into the initial phase of a continuous wave supercontinuum generation seeded by a noise-driven MI.

The defining physics underpinning the development of a breather on a plane wave background is to excite a pair of sidebands. The subsequent dynamics follows a four-wave mixing process and generates cascades of new modulation frequencies in a triangular fashion which are harmonics of the initial pair of MI frequencies. The end product is a full grown breather forming a series of pulses.

However, exciting two pairs of side-bands and their amplification during evolution follows even more complex MI dynamics and cannot be explained by a single breather formulation. Following the principle of nonlinear superposition, two-pair of side band will form a two-breather solution \cite{akhmediev1997solitons, akhmediev1985generation,Kedziora:12:PRE1, Akhmediev:09:PRE} described by the physics of  higher order-modulation instability (H-MI).  Recently, a number of studies have been devoted to study their occurrences in real physical systems \cite{Erkintalo:11:PRL,hammani2011peregrine,erkintalo2012seeded}. Particularly, in \cite{Erkintalo:11:PRL}, the role of H-MI to form a two breather solution in nonlinear fiber optics has been theoretically discussed with experimental verification.

Taking into account of a single pair of side-bands, though the study of MI in nonlinear dispersive system is abound, however, the role of H-MI in connection with nonlinear coherent structures in physical systems is relatively new. The available literature on this topics is also limited. Additionally, the primary features of MI and H-MI mostly studied using the fundamental NLSE which only includes the lowest order dispersion and nonlinear terms. However, the higher-order dispersion and nonlinear terms should be taken into account to improve the accuracy of the equation. In this work, we study the H-MI dynamics in the case when the higher-order terms preserve the integrability of the extended NLSE. We start investigation of the gain properties of H-MI using a fourth-order NLSE given as:
\begin{eqnarray}
\label{tot}
\nonumber  i \psi_x &+& \alpha_2 (\psi_{tt} + 2\psi|\psi|^2)  + \alpha_4 (\psi_{tttt} + 8|\psi|^2\psi_{tt}  \\
& + & 4\psi|\psi_t|^2 + 6\psi_t^2\psi^* + 2\psi^2\psi_{tt}^*+ 6\psi|\psi|^4 )= 0,
\end{eqnarray} 
that have been considered earlier in \cite{porsezian1992integrability,wang2013breather}.
The coefficient $\alpha_2$ in Eq.(\ref{tot}) scales the terms of the standard NLSE while the coefficient $\alpha_4$ scales the fourth-order terms. Without loss of generality, the coefficient $\alpha_2$ can be fixed to $\alpha_2=\frac{1}{2}$. Porsezian {\it at el.} derived the equation in connection with integrability aspects of the one-dimensional Heisenberg spin-chain problem \cite{porsezian1992integrability}. However, physically, we can think of Eq.(\ref{tot}) as a special case of a more general governing equation for pulse propagation in an optical fibre\cite{agrawal2011nonlinear}:
\[
i \psi_{x} =  - \sum_{m=1}^{\infty}\frac{i^m\,\beta_m}{m!} \frac{\partial^m \psi}{\partial t^m }
\]
\begin{equation}
\label{agrw}
-\gamma \left(1+is\frac{\partial }{\partial t }  \right)\, \left(\psi\, \int_{0}^{\infty} R(t')\,|\psi(t-t')|^2  dt' \right),
\end{equation}
This equation is being widely used for ultra-short pulse propagation in optical fiber. The equation is different from Eq.(\ref{agrw}) in the sense that, it is not integrable and cannot be solved analytically. A numerical scheme is required to solve this equation. Where the first term on the right hand side is the expansion of linear dispersion, the second term describes the nonlinear terms. Here, $s$ is the self-steepening coefficient, the coefficient $\gamma$ is defined by the effective core area and $ R(t)$ includes instantaneous and delayed  (Raman) contributions of the nonlinear material response.
The integral in Eq.(\ref{agrw}) is often approximated by taking the series to first-order only, $viz.$ $ |\psi|^2-\tau_R\frac{\partial}{\partial t}(|\psi|^2), $ for Raman delay $\tau_R$. However, in reality, to achieve a greater accuracy, we need higher-order terms, and these involve higher order time derivatives of intensity, $|\psi|^2$. Specifically, $\psi \frac{\partial}{\partial t^2}(|\psi|^2)$ delivers most of the fourth-order terms. In experiments, up to fourth-order terms in this series can be important \cite{hook1993ultrashort}. Thus, the linear fourth-order dispersion term in Eq.(\ref{tot}) describes with $\alpha_4=\beta_4/24$ is balancing the last quintic-nonlinear terms. The other 4 terms describe the nonlinear dispersion.

\section{NLSE-two-breather solution} 
\label{sec44}

Due to MI, a first-order breather solution describe the development of a pair of sidebands as a perturbation on a constant background making a periodic train of pulses \cite{akhmediev1997solitons}. Similarly, two pair of sidebands will also grow and will be developed into a general two-breather solution following the nonlinear superposition mechanism. However, in that case both pair of initial sidebands must remain within the instability band. General expression for a breather solution of any order of Eq.(\ref{tot}) can conveniently given by the expression:
\begin{equation}
\label{lpd-brrr}
\psi_n(x,t)=\left((-1)^n+\frac{G_n+i \,H_n}{D_n}\right) e^{i x}
\end{equation}
where $n$ indicates the order of the solution. The functions $G_n(x,t), H_n(x,t)$ and $D_n(x,t)$ are real. Therefore, the expression for a two breather solution can be given as:
\begin{eqnarray}
\label{vh1}
\nonumber G_2&=&\frac{ \kappa _1^2-\kappa _2^2}{\kappa _1 \kappa _2} \{-\cos \left(t_{\text{s2}} \kappa _2\right) \cosh \left(x_{\text{s1}} \delta _1\right) \delta _2 \kappa _1^3\\
\nonumber&+&\cosh \left(x_{\text{s2}} \delta _2\right) \kappa _2 [\cos \left(t_{\text{s1}} \kappa _1\right) \delta _1 \kappa _2^2\\
\nonumber&+&\cosh \left(x_{\text{s1}} \delta _1\right) \kappa _1 \left(\kappa _1^2-\kappa _2^2\right)]\}
\\
\nonumber H_2&=&\frac{\kappa _1^2-\kappa _2^2}{\kappa _1 \kappa _2}  \{\sinh \left(x_{\text{s2}} \delta _2\right) v_2 [\cos \left(t_{\text{s1}} \kappa _1\right) \delta _1\\
&-&\cosh \left(x_{\text{s1}} \delta _1\right) \kappa _1] \kappa _2^2+\sinh \left(x_{\text{s1}} \delta _1\right) v_1 \kappa _1^2\\
\nonumber&\times& [\cosh \left(x_{\text{s2}} \delta _2\right) \kappa _2-\cos \left(t_{\text{s2}} \kappa _2\right) \delta _2]\}
\\
\nonumber D_2&=&\frac{1}{\kappa _1 \kappa _2} \{\cosh \left(x_{\text{s1}} \delta _1\right) \kappa _1 [\cos \left(t_{\text{s2}} \kappa _2\right) v_2 \kappa _2 \left(\kappa _1^2-\kappa _2^2\right)\\
\nonumber &+&\cosh \left(x_{\text{s2}} \delta _2\right) \kappa _2 \left(-2 \kappa _1^2+\left(-2+\kappa _1^2\right) \kappa _2^2\right)]\\
\nonumber &+&v_1 \kappa _1 [\cos \left(t_{\text{s1}} \kappa _1\right) \cos \left(t_{\text{s2}} \kappa _2\right) \delta _2 \left(\kappa _1^2+\kappa _2^2\right)\\
\nonumber &+&\kappa _2 \{[\sin \left(t_{\text{s1}} \kappa _1\right) \sin \left(t_{\text{s2}} \kappa _2\right)+\sinh \left(x_{\text{s1}} \delta _1\right) \sinh \left(x_{\text{s2}} \delta _2\right)] \\
\nonumber &\times&v_2 \kappa _1 \kappa _2+\cos \left(t_{\text{s1}} \kappa _1\right) \cosh \left(x_{\text{s2}} \delta _2\right) \left(-\kappa _1^2+\kappa _2^2\right)\}]\}.
\end{eqnarray}

With $n=2$, putting $G_2$, $H_2$ and $D_2$ in Eq.(\ref{lpd-brrr}) becomes the second-order breather solution of  the standard NLSE. This solution has been derived and analyzed rigorously in \cite{akhmediev1997solitons,Kedziora:12:PRE1}. 

For a two-breather solution, the two free parameters $\kappa_1$ and $\kappa_2$ are the modulation frequencies for two colliding breathers. These two frequencies separately provide the associated growth rates of each of these breathers, viz. $\delta_1=\frac{1}{2}\kappa_1 \sqrt{4-\kappa_1^2}$ and $\delta_2=\frac{1}{2}\kappa_2 \sqrt{4-\kappa_2^2}$. The parameters $x_{s1}=(x-x_1),x_{s2}=(x-x_2)$ and $t_{s1}=(t-t_1),t_{s2}=(t-t_2)$ are the translations along $x$ and $t$. The modulation frequencies $\kappa_1$ and $\kappa_2$ play central r\^oles in the two-breather solution. During the development process of breathers, $\kappa_1$ and $\kappa_2$ do not act independently, but jointly influence the formation   following a set of rules \cite{chowdury2017breather,Kedziora:12:PRE1}. For example, there are two well-known special cases involving frequencies $\kappa_1$ and $\kappa_2$:

(a) with $\kappa_1=\kappa_2=\kappa$,  Eq.(\ref{vh1}) reduces to a degenerate second-order breather solution  \cite{akhmediev1997solitons,Kedziora:12:PRE1}.
 
 (b) with $\kappa \to 0$ in the above case, Eq.(\ref{vh1}) reduces to a second-order rational solution  \cite{akhmediev1997solitons,Kedziora:12:PRE1}.
 
 These choice of MI frequency is equally applicable to all the higher-order solutions as well. However, recent research has revealed that the role of $\kappa_1$ and $\kappa_2$ even for a second-order NLSE solution is far more complicated and plays a wider r\^ole in producing  a whole new classes of solutions.  A series of recent works has been devoted to this theme \cite{Ankiewicz:11:PLA,Kedziora:11:PRE,Kedziora:12:PRE2,Kedziora:13:PRE}. 

\begin{figure}[htbp]
\begin{center}
\includegraphics[width=0.3\textwidth]{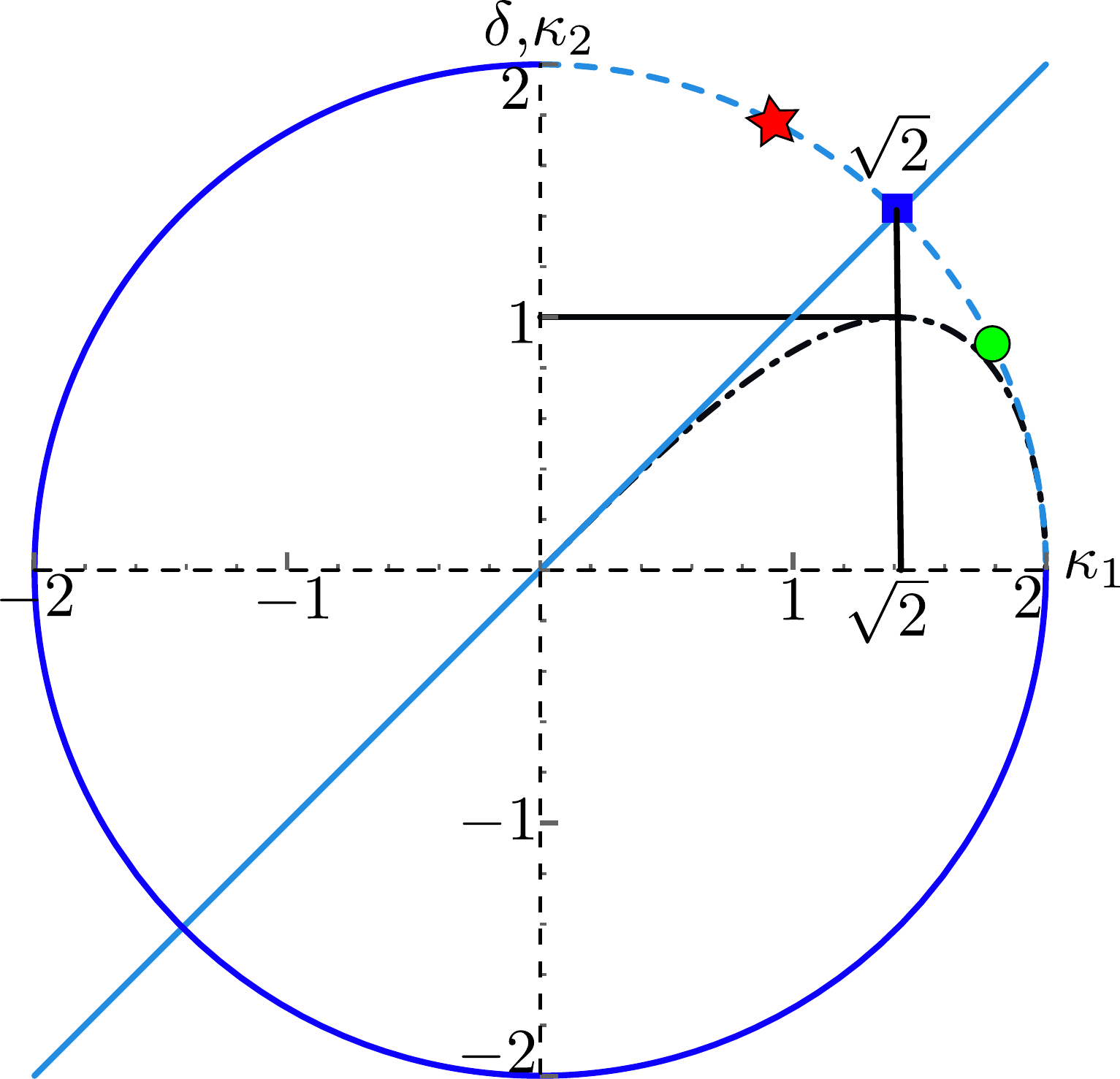}
\caption{(Color online) Plot of the solutions of Eq.(\ref{gr}) $\kappa_2=\kappa_1$ and $\kappa_2=\pm \sqrt{4-\kappa_1^2}$. Growth is equal on both the  {\it solid light blue} line and the {\it dashed light blue} curve. Both of them intersects at $\sqrt{2}$.  {\it Black dot-dashed} curve is the NLSE MI band. The black vertical line passes through the maximum growth point of NLSE first-order breather MI band intersect the {\it dashed light blue} curve at $\sqrt{2}$ where a two breather solution of NLSE is undefined.  }
\label{Fig-Circle}
\end{center}
\end{figure}

\begin{figure}[htbp]
\label{fig2}
\begin{center}
\subfloat[]{\label{Fig-1}\includegraphics[width=0.23\textwidth]{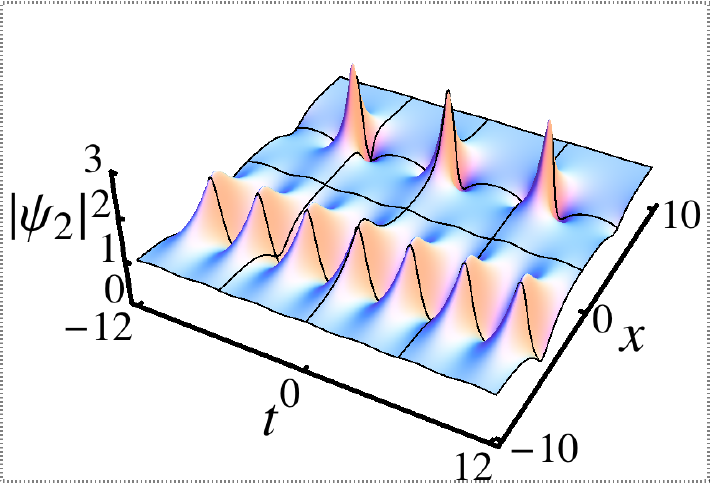}}
\subfloat[]{\label{Fig-2}\includegraphics[width=0.23\textwidth]{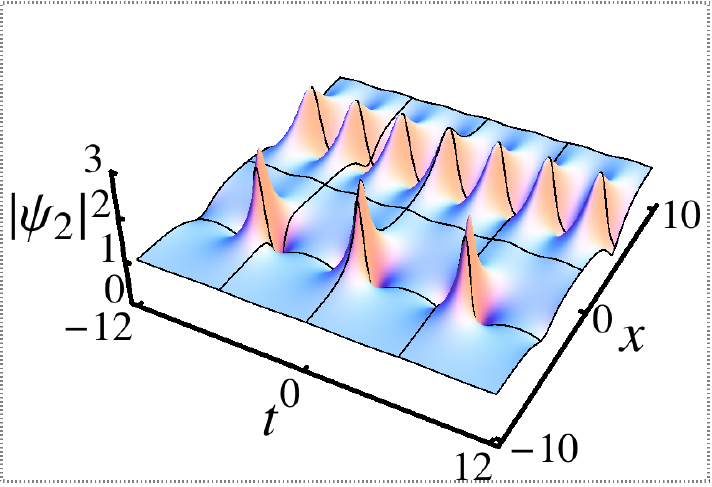}}
\\
\subfloat[]{\label{Fig-3}\includegraphics[width=0.23\textwidth]{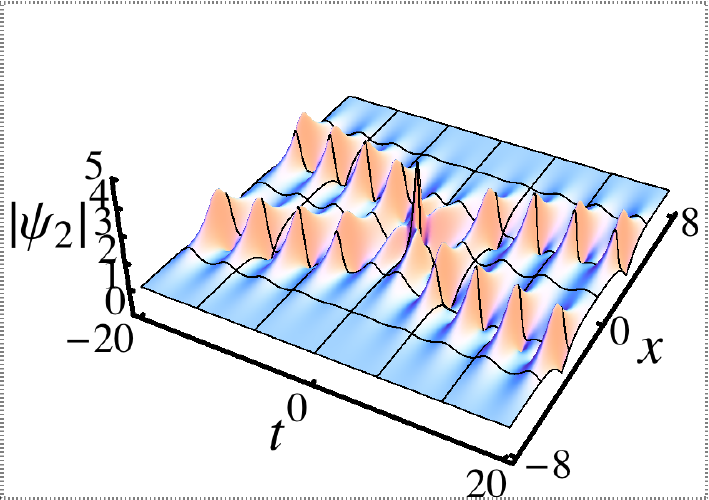}}
\caption{\label{Fig:second-breather}(Color online)   (a) Second-order NLSE breather solution with $\kappa_1=\frac{2}{\sqrt{5}}$, $t_1=0$ and $t_2=\pi/2\kappa_1$ the red star in Fig.(\ref{Fig-Circle}) ; while (b) with $\kappa_1=\frac{4}{\sqrt{5}}$ and $t_1=\pi/\kappa_1$ and $t_2=0$, the green small circle, for (c) $\kappa=\sqrt{2}$, the blue small rectangle (should I put the solution?), for both (a) and (b), translation $x_1=5$, $x_2=-5$.}
\end{center}
\end{figure}

\begin{figure}[htbp]
\begin{center}
\includegraphics[width=0.50\textwidth]{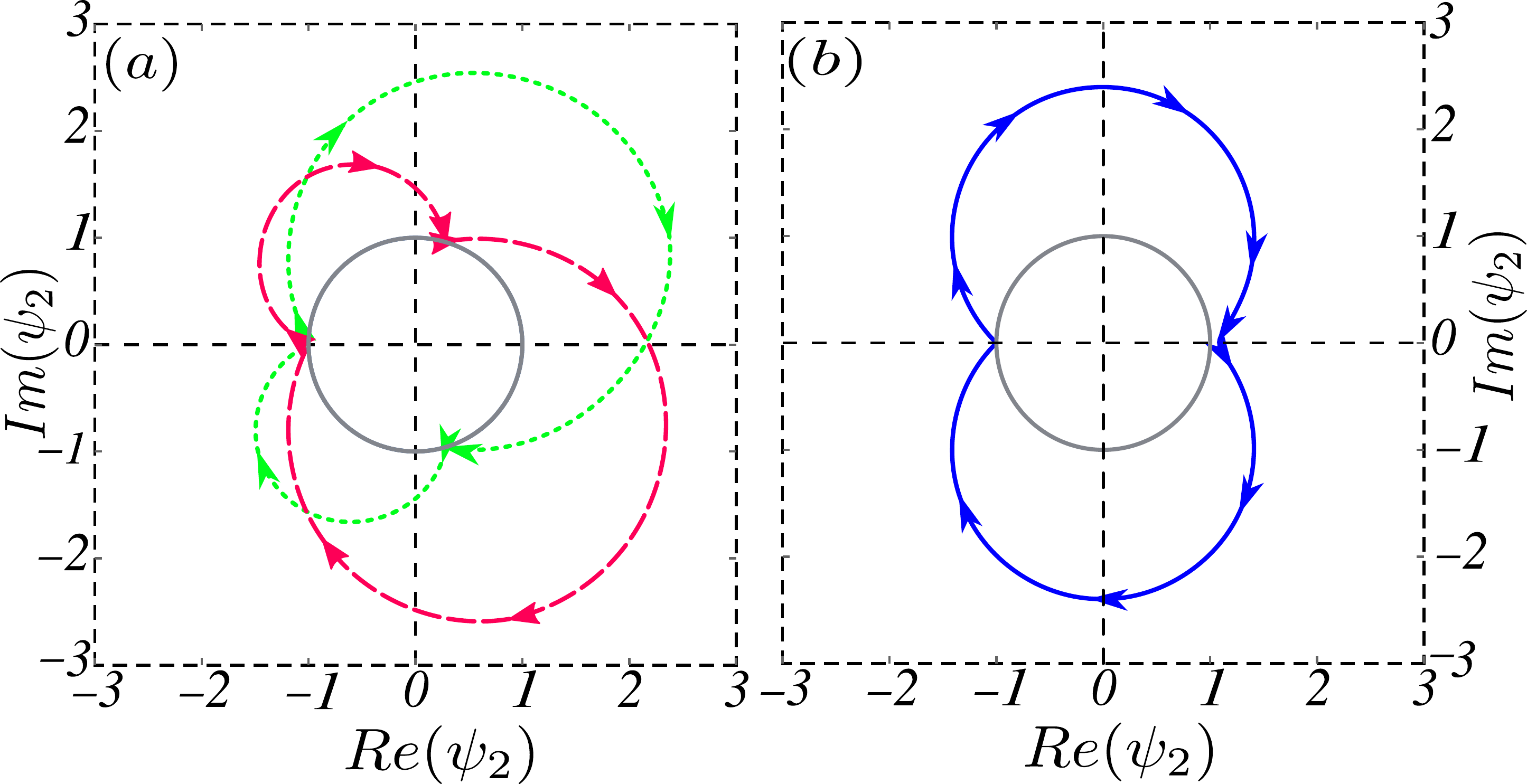}
\caption{\label{Fig:second-breather}(Color online)   (a) Second-order NLSE breather solution on a complex-plane. Parameter values are same as Fig.(\ref{Fig-1},\ref{Fig-2}) (b) second-order degenerate NLSE breather solution with $\kappa=\sqrt{2}$ corresponds to Fig.(\ref{Fig-3}) .}
\label{fig3}
\end{center}
\end{figure}

In previous works, following the first case mentioned in (a) above \cite{akhmediev1997solitons,Kedziora:12:PRE1}, the choice of setting $\kappa_1=\kappa_2$ has been considered somewhat arbitrarily as a condition for an equal frequency scenario. The main feature of this solution is, even it is a two breather solution, controlled by only one frequency parameter having same growth for each of the breather in the solution. 

However, in the present case, investigation reveals that  this special relation ($\kappa_1=\kappa_2$) can be derived from a simple mathematical relation. Specifically, we can obtain  $\kappa_1=\kappa_2$ as one of the solutions of the equation:
\begin{equation}
\label{gr}
\delta_1-\delta_2=0
\end{equation}
 Let us begin a systematic study:  if we put $\delta_1=\delta_2$ in Eq.(\ref{vh1}) and solve it for $\kappa_2$, we find three real solutions, $\kappa_2=\kappa_1$ and $\kappa_2=\pm \sqrt{4-\kappa_1^2}$. We plot these solutions in Fig.({\ref{Fig-Circle}}). In  practice, on the entire parameter space of $\kappa_1$ and $\kappa_2$, these solutions  give us the specific locus where the growth-rates of the breathers are equal. The first solution, $\kappa_1=\kappa_2$ is the so-called degenerate solution condition where the solution becomes undefined. However, using l'H\^ospital's rule, the solution can be recovered even in this case \cite{akhmediev1997solitons,Kedziora:12:PRE1}. 

This equal-growth, degenerate solution is seen in   Fig.({\ref{Fig-Circle}}) by following the blue-solid line passing through the origin. It intersects the circle at $\sqrt{2}$ indicated by the small blue rectangle. Every point on this line represents $\kappa_1=\kappa_2$ with growth $\delta_1=\delta_2$.  Hence, when $\kappa_1 \to \kappa_2=\kappa$, following l'H\^ospital's rule, it becomes the degenerate breather solution. With $\kappa \to 0$, following the blue line to the origin, the solution becomes a second-order rational solution. Both of these solutions have been studied and presented in \cite{akhmediev1997solitons,Kedziora:12:PRE1}. 

However, the r\^ole played by the solutions $\kappa_2=\pm \sqrt{4-\kappa_1^2}$  in the second order NLSE solution Eq.(\ref{vh1}) is the current focus of our investigation. Specifically, we present here, for the first time,  the solution $\kappa_2=\pm \sqrt{4-\kappa_1^2}$ as a condition of $\delta_1=\delta_2$ and its subsequent analysis. It is interesting to note that the condition $\delta_1=\delta_2$, in principle reveals the entire geometric interpretation of the parameter space $\kappa_1$ and $\kappa_2$. For example, from the solution 
$\kappa_2=\pm \sqrt{4-\kappa_1^2}$, we can say that the parameter space is bounded by the circle:
\begin{equation}
\label{bound}
\kappa_1^2+\kappa_2^2=4
\end{equation}
with radius $2$, as depicted in Fig.({\ref{Fig-Circle}}). On this circle, we deal only with the first quadrant identified by the {\it blue dashed} curve, as $\kappa_1$ and $\delta_1$ remain positive and real here. In that case we use the solution $\kappa_2=\sqrt{4-\kappa_1^2}$. Apart from these loci, one can choose $\kappa_1$ and $\kappa_2$ arbitrarily, while the frequency range remains within the MI band. However, if we use $\kappa_2=\sqrt{4-\kappa_1^2}$ in Eq.(\ref{vh1}), we have the expression:
\begin{eqnarray}
\label{vh2}
\nonumber G_2&=&\left(2-\kappa _1^2\right) \{-\cos \left(t_{\text{s1}} \kappa _1\right) \cosh \left(x_{\text{s2}} \delta _1\right) \left(4-\kappa _1^2\right){}^{\frac{3}{2}}\\
\nonumber &+&\cosh \left(x_{\text{s1}} \delta _1\right) [\cos \left(t_{\text{s2}} v_1\right) \kappa _1^3\\
\nonumber &-&4 \cosh \left(x_{\text{s2}} \delta _1\right) \left(-2+\kappa _1^2\right)]\}
\\
\nonumber H_2&=&-\kappa _1 \left(-2+\kappa _1^2\right) \{\sinh \left(x_{\text{s1}} \delta _1\right) v_1 [-2 \cosh \left(x_{\text{s2}} \delta _1\right)\\
\nonumber &+&\cos \left(t_{\text{s2}} v_1\right) \kappa _1]+\sinh \left(x_{\text{s2}} \delta _1\right) [2 \cosh \left(x_{\text{s1}} \delta _1\right) v_1\\
 &+&\cos \left(t_{\text{s1}} \kappa _1\right) \left(-4+\kappa _1^2\right)]\}
\\
\nonumber D_2&=&\frac{1}{v_1} \{\left(-4+\kappa _1^2\right) \{-\kappa _1 \{2 \cos \left(t_{\text{s2}} v_1\right) \cos \left(t_{\text{s1}} \kappa _1\right)\\
\nonumber &+&[\sin \left(t_{\text{s2}} v_1\right) \sin \left(t_{\text{s1}} \kappa _1\right)\\
\nonumber &+&\sinh \left(x_{\text{s1}} \delta _1\right) \sinh \left(x_{\text{s2}} \delta _1\right)] v_1 \kappa _1\}\\
\nonumber &+&2 \cos \left(t_{\text{s1}} \kappa _1\right) \cosh \left(x_{\text{s2}} \delta _1\right) \left(-2+\kappa _1^2\right)\}\\
\nonumber &+&\cosh \left(x_{\text{s1}} \delta _1\right) v_1 [2 \cos \left(t_{\text{s2}} v_1\right) \kappa _1 \left(-2+\kappa _1^2\right)\\
\nonumber &-&\cosh \left(x_{\text{s2}} \delta _1\right) \left(8-4 \kappa _1^2+\kappa _1^4\right)]\}
\end{eqnarray}
 Clearly, this is a two breather solution with a common growth parameter $\delta_1$ and frequency $\kappa_1$ and it posses several interesting properties. Generally speaking, the parameter space for $\kappa_1$ and $\kappa_2$ are now can be identified by the area of the circle (Fig.({\ref{Fig-Circle}})) defined by the solution $\kappa_2=\pm \sqrt{4-\kappa_1^2}$. The growth rate is equal on the circumference and the line passing through the origin. One of the interesting feature of the solution is, it demonstrate a clear correlation between the growth rate and the amplitude of the constituent breathers. For example, even the growth is equal, the breathers may have different amplitude presented by Fig.(\ref{Fig-1},\ref{Fig-2}). The corresponding $\kappa_1$ values are given by the {\it red star} and the {\it green circle}. Their complex plane representation is given in Fig.(\ref{fig3}a). There is no other place except $\kappa_2=\pm \sqrt{4-\kappa_1^2}$, where a breather will have different amplitude but with same growth $\delta_1$. The breather will have equal amplitude with same growth rate only on the straight line passing through the origin.

Another feature of the solution Eq.(\ref{vh2}) is, for the degenerate solution where $\kappa_2=\kappa_1$, and with the limit, $\kappa_1 \to 0$, while we get a second-order rogue wave solution  \cite{akhmediev1997solitons,Kedziora:12:PRE1}, with $\kappa_2=\sqrt{4-\kappa_1^2}$, (Eq.(\ref{vh2})), we have the freedom to take two such limits: one is $\kappa_1 \to 0$ and the other is $\kappa_1 \to 2$. For each limit, we obtain a first-order rogue wave solution.  Alternatively, if we take these limit on Eq.(\ref{vh2}), we get:
\begin{eqnarray}
\label{two-limit}
\nonumber \lim_{\kappa_1 \to 0} \psi_2(x,t)&=&\lim_{\kappa_1 \to 2} \psi_2(x,t)\\ 
 &=& \psi_1(x,t)=\left(\frac{G_1+i \,H_1}{D_1}-1\right) e^{i x}
\end{eqnarray}
where $G_1$, $H_1$ and $D_1$ can be given as:
\begin{eqnarray*}
\label{nlse-r-11}
  \nonumber G_1&=&4, \\ \nonumber
  H_1&=& 8x_{s1} ,  \\
  D_1&=&1+4t_{s1}^2+4 x_{s1}^2
\end{eqnarray*} 
 The square {\it red} dot corresponds to the point where $\kappa_1=\kappa_2=\sqrt{2}$ and the solid {\it blue} line and on the {\it yellow} curve cross  each-other.  The solution is undefined at this point. Note that, at this same point on the MI band, an NLSE breather has its maximum growth.
 
\section{Two-breather LPDE}
\label{sec5}

Two breather solution of LPDE is derived based on two eigenvalues, $\lambda_j$ with $j=1,2$. They are expressed in terms of modulation frequencies of the two breathers, $\kappa_j=2\sqrt{1+\lambda_j^2}$ . The derivation technique named as {\it Darboux transformation} that we employ to derive the solution is described in \cite{akhmediev1997solitons}. With $n=2$, in the general expression (\ref{lpd-brrr}), the two-breather solution of Eq.(\ref{tot}) is given by:
\begin{eqnarray}
\label{lpd-br2}
\nonumber G_2&=&\frac{\kappa _1^2-\kappa _2^2}{\kappa _1 \kappa _2} \big\{- \delta _2 \kappa _1^3\cos \left(t_{\text{s2}} \kappa _2\right) \cosh \left(V_{\text{H1}} x_{\text{s1}}\right)\\
\nonumber&+&\kappa _2\cosh \left(V_{\text{H2}} x_{\text{s2}}\right)  [ \delta _1 \kappa _2^2\cos \left(t_{\text{s1}} \kappa _1\right)\\
\nonumber&+& \kappa _1\left(\kappa _1^2-\kappa _2^2\right)\cosh \left(V_{\text{H1}} x_{\text{s1}}\right) ]\big\}\\
\nonumber H_2&=&\frac{2 \left(\kappa _1^2-\kappa _2^2\right)}{\kappa _1 \kappa _2}  \big\{ \delta _2  \kappa _2 \sinh \left(V_{\text{H2}} x_{\text{s2}}\right)[\delta _1 \cos \left(t_{\text{s1}} \kappa _1\right) \\
\nonumber&-& \kappa _1\cosh \left(V_{\text{H1}} x_{\text{s1}}\right)]+ \delta _1 \kappa _1\sinh \left(V_{\text{H1}} x_{\text{s1}}\right) \\
\nonumber&\times&[\kappa _2 \cosh \left(V_{\text{H2}} x_{\text{s2}}\right)-\delta _2\cos \left(t_{\text{s2}} \kappa _2\right) ]\big\}\\
\nonumber D_1&=&\frac{1}{\kappa _1 \kappa _2} \big\{\kappa _1 \cosh \left(V_{\text{H1}} x_{\text{s1}}\right)\{2 \delta _2 \left(\kappa _1^2-\kappa _2^2\right) \cos \left(t_{\text{s2}} \kappa _2\right)\\
\nonumber&+&\kappa _2\cosh \left(V_{\text{H2}} x_{\text{s2}}\right)[\kappa _1^2 \left(\kappa _2^2-2\right)-2 \kappa _2^2] \}\\
\nonumber&+&2 \delta _1 \big[\delta _2 \left(\kappa _1^2+\kappa _2^2\right) \cos \left(t_{\text{s1}} \kappa _1\right) \cos \left(t_{\text{s2}} \kappa _2\right)\\
\nonumber&+&\kappa _2 \big(2 \delta _2 \kappa _1[\sin \left(t_{\text{s1}} \kappa _1\right) \sin \left(t_{\text{s2}} \kappa _2\right)\\
&+&\sinh \left(V_{\text{H1}} x_{\text{s1}}\right) \sinh \left(V_{\text{H2}} x_{\text{s2}}\right)]\\
\nonumber&+&\left(\kappa _2^2-\kappa _1^2\right) \cos \left(t_{\text{s1}} \kappa _1\right) \cosh \left(V_{\text{H2}} x_{\text{s2}}\right)\big)\big]\big\}
\end{eqnarray}
where $\delta_j=\kappa_j \sqrt{4-\kappa_j^2}/2$, $V_{Hj}=2\delta_j\left[\alpha_2-\alpha_4(\kappa_j^2-6)\right]$, $x_{sj}=(x-x_j)$ and $t_{sj}=(t-t_j)$ with $x_j$ and $t_j$ being translations along $x$ and $t$ axes. Modulation frequencies $\kappa_1$ and $\kappa_2$ are two independent free parameters that play   key roles in  the mutual interaction between the breathers. Depending on the parameter  $\kappa_j$,  Eq.(\ref{lpd-br2}) may represent either the superposition two fundamental AB or KM solitons \cite{akhmediev1997solitons}. When $\alpha_4=0$, the solution reduces to the standard NLSE two breather solution Eq.(\ref{vh1}) presented in \cite{akhmediev1985generation,Kedziora:12:PRE1}. For convenience, growth-rates of Eq.(\ref{lpd-br2}) can be given as:
\begin{eqnarray}
 V_{H1} & =  & 2\delta_1\left[\alpha_2-\alpha_4(\kappa_1^2-6)\right]\\
V_{H2} & =  & 2\delta_2\left[\alpha_2-\alpha_4(\kappa_2^2-6)\right] \\
\nonumber \end{eqnarray}
Two-breather solutions of Eq.(\ref{tot}) in various forms have been studied in \cite{wang2016breather,yang2017breathers}. However, the results were incomplete and require more detailed analysis. Several important special cases were missing 
and we investigate them here. When we go beyond the simple first-order equation, the number of parameters also increases and the dynamics of two-breather interactions becomes more complex. 
In an effort to understand these complexities, taking several parameters to be equal sometimes reveals explicit features of the breather dynamics, such as their collisions, superpositions and transformations \cite{chowdury2015breather,chowdury2015breathertosoliton,chowdury2015moving,ankiewicz2016superposition}.

Our present work is another step forward in understanding the complex higher-order MI dynamics that is involved with the two-breather solutions of extended NLSE equations. Similar to the NLSE two-breather solution in Section \ref{sec44}. Here we equate the modulation parameters $V_{H1}$ and $V_{H2}$ and solve the equation:
\begin{equation}
\label{lpde}
V_{H1}-V_{H2}=0
\end{equation}
for $\kappa_2$ with arbitrary $\alpha_4$ where $\alpha_2=1/2$.  Equation (\ref{lpde}) is an {\it octic} that is an eighth degree polynomial in $\kappa_j$. We obtain eight solutions, where two are $\kappa_2=\pm \kappa_1$ and we provide the other six solutions, $a_{12}, a_{34}$ and $a_{56}$ in the Appendix \ref{appen}.
 
These solutions will give us the loci for equal growth revealing the whole scenarios of parameter space and allow us to provide a comprehensive and explicit description of MI for a two-breather solution of LPDE.  If we use these solutions as values of $\kappa_2$ in Eq.(\ref{lpd-br2}), we will have a solution with only one frequency, $\kappa_1$, and one growth rate, $V_{H1}$, similar to   Eq.(\ref{vh2}). With the solution $\kappa_1=\kappa_2$, we have the degenerate solution of the LPDE that already given in \cite{chowdury2017breather}. With the above solutions, now we have two scenarios to consider, viz. (a) $\alpha_4>0$ and (b) $\alpha_4<0$. 

\subsection {\it Case (a) $\alpha_4>0:$} With $\alpha_4>0$, in Eq.(\ref{polyp}), we will have four real solutions; first two are $\kappa_1=\pm \kappa_2$ and the other two are $a_{56}$ given in Appendix (\ref{appen}) and plotted in Fig.(\ref{Fig-apositive}). For example, with $\alpha_4 <<1$ say, for $\alpha_4=1/128$ which is near to zero, we will have the same circular solution as for the basic NLSE, indicated by the {\it blue solid} circle. The {\it blue solid} line passing through the origin is the  $\kappa_1=\pm \kappa_2$ line where solution remains degenerate and also the associated MI band remains similar to the basic NLSE MI band. For positive value of $\alpha_4$, the Fig.(\ref{Fig-apositive}) represents the parameter space $\kappa_1$ and $\kappa_2$ for solution Eq.(\ref{lpd-br2}). We also put the growth rate in the plot to illustrate how the parameter space changes with changing MI band. For real growth rate, we focus on the first quadrant of the plot. 

\begin{figure}[htbp]
\begin{center}
\includegraphics[width=0.3\textwidth]{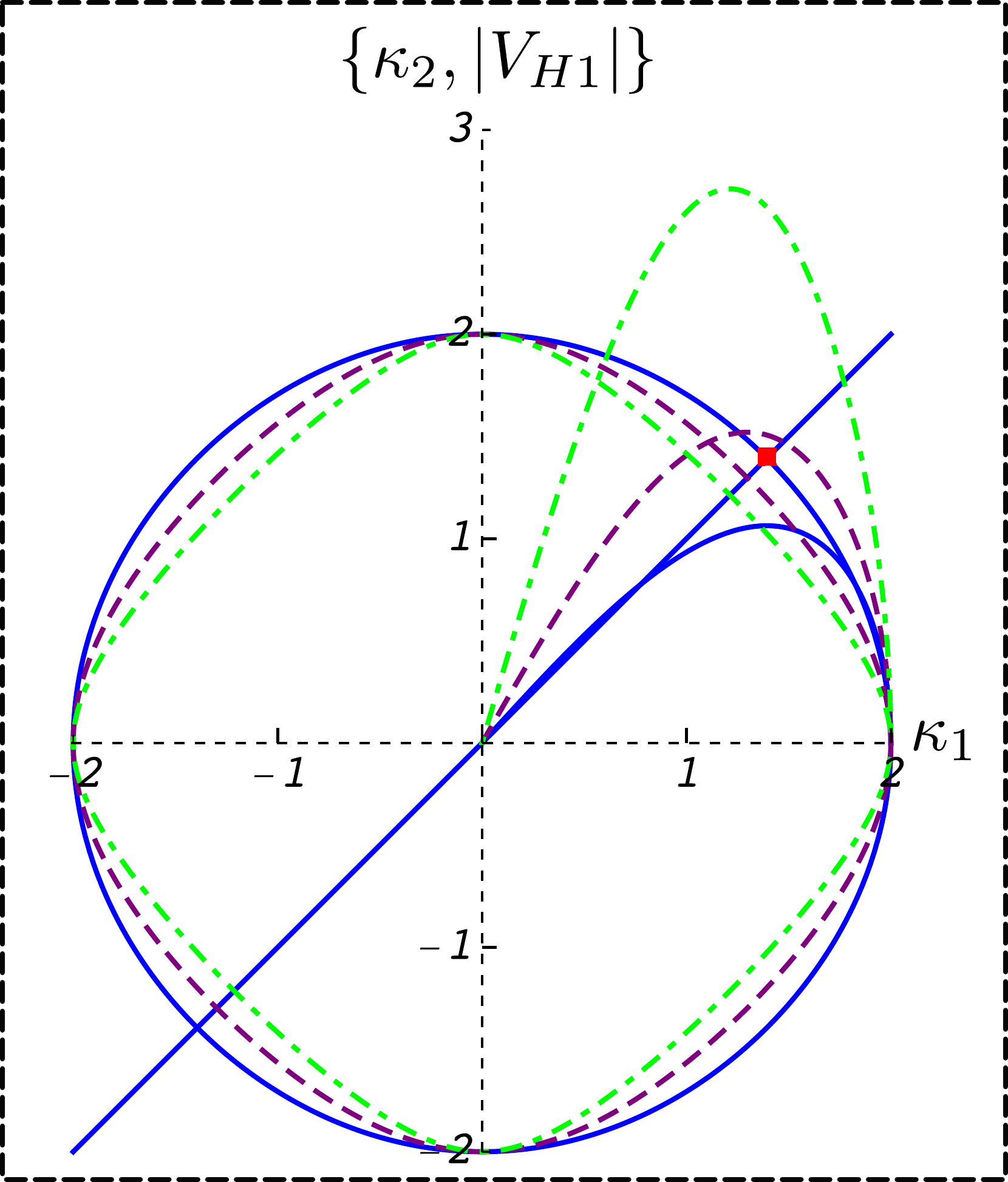}
\caption{(Color online) Locus of equal growth rate and  parameter space for $\kappa_1$ and $\kappa_2$ for LPDE two-breather solution. The {\it blue-circle} and the growth rate $V_{H1}$ is for $\alpha_4=1/128$, the {\it purple-dashed line} is for $\alpha_4=1/16$ and the $\it green\, dot\,dashed\, line$ is for $\alpha_4=1/5$. Associated MI curve is also presented with same colour profile, respectively. For all three case $\alpha_2=1/2$.}
\label{Fig-apositive}
\end{center}
\end{figure}

In general, positive value of $\alpha_4$ reinforces the focusing effects making a faster gain. As we increase the value of $\alpha_4$, the circular parameter space defined for the basic NLSE now becomes distorted. For example, with $\alpha_4=1/16$, we have the distorted circle with the {\it purple-dashed} line and the associated MI band with the ({\it purple-dashed} MI curve), now with raised height due to the increased strength of focusing effects. Gradually, the distortion will continue with the increased value of $\alpha_4$ while raising the height of MI band. Finally, with $\alpha_4=1/5$, we have the {\it green dot-dashed} line and its associated MI band raised significantly. Nevertheless, with $\alpha_4 \simeq 1$, the circular solution gradually becomes a square in shape.

\subsection {\it Case (b) $\alpha_4<0:$}
\label{positive}
\begin{figure}[htbp]
\begin{center}
\includegraphics[width=0.3\textwidth]{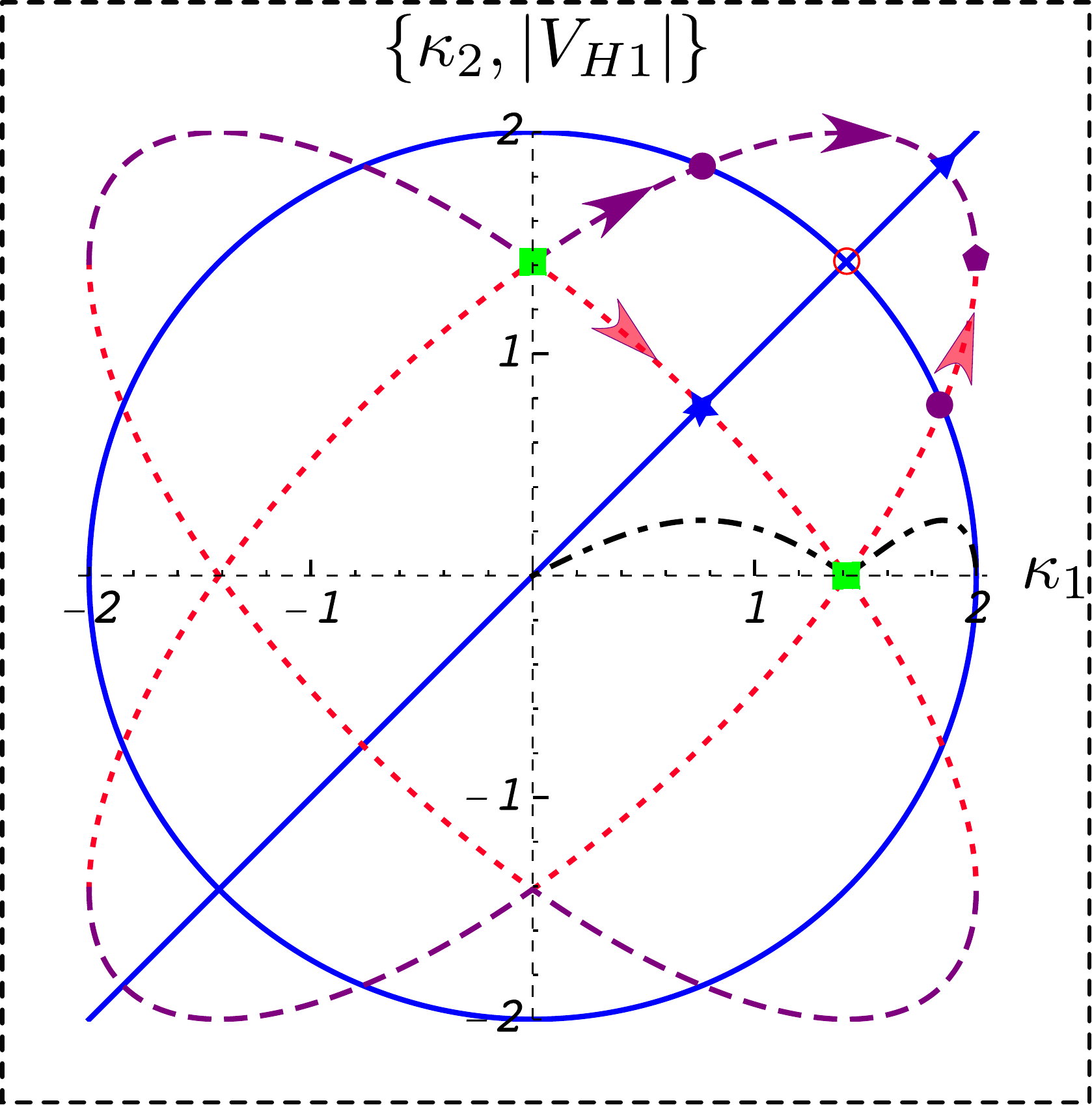}
\caption{(Color online) Plot of the locus presented by the solutions $a_{12}'$ which is the {\it blue circle} and both $a_{34}'$ and $a_{56}'$ together making the two intersecting ellipses with {\it red dotted line} and {\it purple dashed line} given in Eq.(\ref{poly}). For all these three intersecting circles and ellipses, the single parameter value is $\alpha_4=-1/8$ and $\kappa_1$ is from $-2 $ to $2$. The {\it black\, dot\,dashed line} which is the growth rate $V_{H1}$, the associated MI curve for $\alpha_4=-1/8$ touched at $\kappa_2=0$ line at $\sqrt{2}$.}
\label{Fig-negative}
\end{center}
\end{figure}

The scenario becomes increasingly complex when we takes negative $\alpha_4$. One of the representative example is $\alpha_4=-1/8$ for which we have a simplified polynomial equation from Eq.(\ref{lpde}). Remarkably, at this particular value of $\alpha_4$, we have equal gain. If we directly put $\alpha_4=-1/8$ in Eq.(\ref{lpde}), we have the simplified polynomial equation as:
\begin{eqnarray}
\label{simplified}
(\kappa_1^2-\kappa_2^2)\\
&\times&(\kappa_1^2+\kappa_2^2-4)\{\kappa_1^4+\kappa_2^4-4(\kappa_1^2+\kappa_2^2)+4\}=0\nonumber
\end{eqnarray}
This is a simplified {\it octic} equation. Clearly, first part of the equation have the solution $\kappa_2=\pm \kappa_1$ and the middle part is a regular circle. Finally, the last part of the equation is two intersecting ellipses. If we solve the equation for $\kappa_2$, including $\kappa_2=\pm \kappa_1$, we have eight closed form real solutions; these can be given as:
\begin{eqnarray}
 \label{poly}
\nonumber a_{12}'&=&\pm (4-\kappa _1^2)^{1/2}
\\
 a_{34}'&=&\pm \left[2-\sqrt{\kappa _1^2 \left(4-\kappa _1^2\right)}\right]^{1/2}
\\
\nonumber a_{56}'&=&\pm \left[2+\sqrt{\kappa _1^2 \left(4-\kappa _1^2\right)}\right]^{1/2}
\end{eqnarray}
These solutions are individually part of an intersecting circle and two ellipses. Compared to the NLSE parameter space in Fig.(\ref{Fig-Circle}), for the LPDE, two new ellipses are now introduced and MI scenarios for equal growth follow more paths than their NLSE counterparts. In the Fig.(\ref{Fig-negative}), we plot these eight solutions and also include  the associated growth rate, $V_{H1}$ in the plot to show how the parameter space re-distributes with the growth-curve.  

It is important to note here that, while these intersecting circle and ellipses define the parameter space of MI frequency $\kappa_1$ and $\kappa_2$ of Eq.(\ref{lpd-br2}), the growth-rate of the breathers remain equal on the boundary. From Fig.(\ref{Fig-negative}), it is now possible to explicitly determine the dynamics of the breather Eq.(\ref{lpd-br2}). For example, the solution $a_{12}'$ is a circle, and we dealt this in Section (\ref{sec44}) for the basic NLSE. However, if we put $a_{12}'$ in Eq.(\ref{lpd-br2}), similarly to   Eq.(\ref{vh2}), we will have an LPDE equal growth solution with only one frequency component, $\kappa_1$ and a growth rate $V_{H1}$. This allow us to have two limits, one at $\kappa_1 \to 0$ and the other at $\kappa_1 \to 2$, and, in both of these limits, we get an LPDE first-order rogue wave solution given by:
\begin{eqnarray}
\label{lpd-r-1}
\nonumber G_1&=&4, \\ \nonumber
H_1&=&16 Bx_{s1} ,  \\
D_1&=&1+4t_{s1}^2+16 B^2 x_{s1}^2
\end{eqnarray} while $\omega = 2(\alpha_2+3\alpha_4)$.
Here the stretching factor $B=(\alpha_2+6\alpha_4)$. If $\alpha_4=0$, the solution Eq.(\ref{lpd-r-1}) reduces to the standard rogue wave solution of the NLSE Eq.(\ref{two-limit}) \cite{Akhmediev:09:PRE}. In Fig.(\ref{Fig-negative}), the solution $a_{34}'$, presented by the {\it red dotted} line, come in contact with the $y$ axes at $(\kappa_1,\kappa_2)=(0,\sqrt{2})$ in the first quadrant and intersects at $x$ axes at $(\kappa_1,\kappa_2)=(\sqrt{2},0)$. Both of these points are identified as {\it green rectangle} on the $x$ and $y$ axes. We may investigate the structure of the breather at these points. To do this, first we put $a_{34}'$ as $\kappa_2$ value in Eq.(\ref{lpd-br2}) that makes the equation defined by a single MI frequency $\kappa_1$ with one gain parameter $V_{H1}$. If we assign $\kappa_1=\sqrt{2}$ that is the {\it green rectangle} on the $x$ makes the solution undefined as the breather corresponds to $\kappa_2$ becomes zero. Similarly, if we put $\kappa_1=0$, again the solution becomes undefined. However, we can recover the solution either taking the limit $\kappa_1 \to 0$ or $\kappa_1 \to \sqrt{2}$ in Eq.(\ref{lpd-br2}). In both of these limits we get a second-order periodic solution:
\begin{eqnarray}
\label{lpd-br22}
\nonumber G_2&=&-2 \left\{-3+2 \sqrt{2} \cos \left(\sqrt{2} t_{\text{s2}}\right)+4 t_{\text{s1}}^2+x_{\text{s1}}^2\right\}\\
 H_2&=&4 \left\{-2+\sqrt{2} \cos \left(\sqrt{2} t_{\text{s2}}\right)\right\} x_{\text{s1}}\\
\nonumber D_2&=&\sqrt{2} \cos \left(\sqrt{2} t_{\text{s2}}\right) \left(-7+4 t_{\text{s1}}^2+x_{\text{s1}}^2\right)\\
\nonumber &+&2 \left\{5+4 t_{\text{s1}} \left[-2 \sin \left(\sqrt{2} t_{\text{s2}}\right)+t_{\text{s1}}\right]+x_{\text{s1}}^2\right\}
\end{eqnarray}
with $\omega=\exp (ix/4)$ in Eq.(\ref{lpd-brrr}). Example of this solution is shown in Fig.(\ref{Fig-negative-sol}). 

We can give a brief description of the evolution of the breather Eq.(\ref{lpd-br2}) following the locus $a_{34}'$ which has two parts. 
\begin{itemize}
\item We define the one end of the first part at the point $(\kappa_1,\kappa_2)=(0,\sqrt{2})$, the first {\it green-rectangle} where the solution is a second order periodic Eq.(\ref{lpd-br22}). 
\item Along the {\it red dotted line} as $\kappa_1$ increases, $\kappa_2$ decreases and we have a two breather solution at these frequencies where growth is equal but with different amplitude until it intersects the degenerate line $\kappa_1=\kappa_2$ which is the point $(\kappa_1,\kappa_2)=(\sqrt{2-\sqrt{2}},\sqrt{2-\sqrt{2}})$ presented by the {\it blue star}. At this point, the solution becomes undefined. But using  l'H\^ospital's rule, the solution can be recovered. Only any points on this line as MI frequency will deliver a two breather where both gain and amplitude is equal. 
\item After crossing the degenerate line, $a_{34}'$ meet at $(\kappa_1,\kappa_2)=(\sqrt{2},0)$, presented by the second {\it green-rectangle} which is an inflection point. At this point, the breather becomes the periodic solution, Eq.(\ref{lpd-br22}), again. This is the first part of the {\it red-dotted} line. 

\item The second part of the {\it red-dotted} line marked by two ends, one at the end points of the first part that is $(\kappa_1,\kappa_2)=(\sqrt{2},0)$ and the other end is upward direction reached at $(\kappa_1,\kappa_2)=(2,\sqrt{2})$. 
\item But before it reached to the upper end it intersects the {\it blue-circle} at presented by the solution $a_{12}'$ at $(\kappa_1,\kappa_2)=(\sqrt{2+\sqrt{2}}$ indicated by the {\it purple solid dot }. Two breather solution remain defined at this point with different amplitude at same gain. 
\item Along this line ({\it red-dotted} line), as $\kappa_1 \to 2$, $\kappa_2 \to \sqrt{2}$ thus the point $(\kappa_1,\kappa_2)=(2,\sqrt{2})$ identified as the {\it purple-polygon}. However, at this point solution Eq.(\ref{lpd-br2}) still remain valid and becomes:
\begin{equation}
\label{periodic}
\psi_1=\frac{e^{\frac{i x}{4}} \cos \left(\sqrt{2} t_{\text{s2}}\right)}{\sqrt{2}-\cos \left(\sqrt{2} t_{\text{s2}}\right)}
\end{equation}
with $\kappa_1=2$. This is a first-order periodic solution and is graphically same as Fig.(\ref{Fig-rogue2}). 
\end{itemize}

\begin{figure}[htp]
\begin{center}
\subfloat[]{\label{Fig-1}\includegraphics[width=0.23\textwidth]{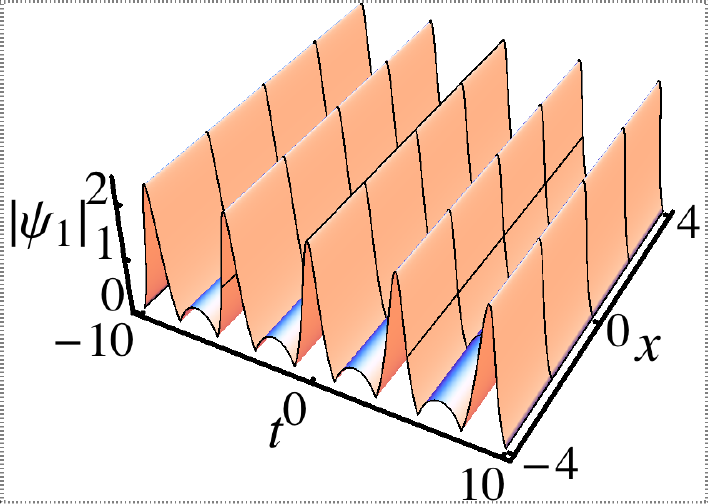}}
\subfloat[]{\label{Fig-2}\includegraphics[width=0.23\textwidth]{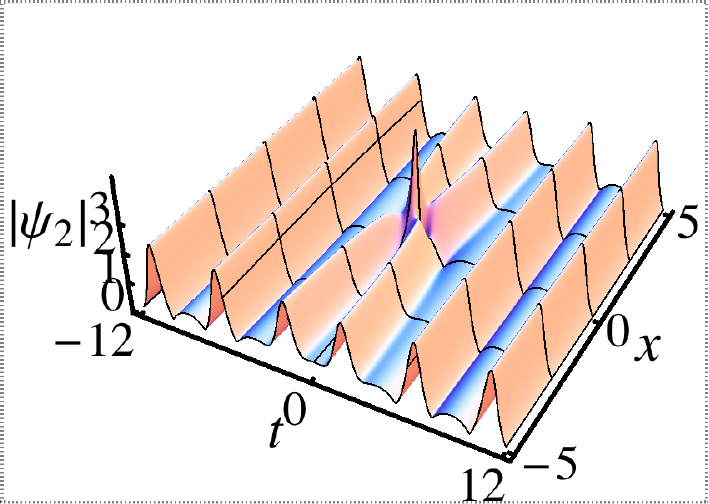}}
\caption{(Color online) A (a) first and (b) second-order periodic solution Eq.(\ref{periodic}) and (\ref{lpd-br22}).
}
\label{periodic-solns}
\end{center}
\end{figure}

A similar description also applies to the solution $a_{56}'$ presented by the {\it purple-dashed} line in Fig.(\ref{Fig-negative}). 

\begin{itemize}
\item In the first quadrant, we define the starting point of the solution $a_{56}'$ at $(\kappa_1,\kappa_2)=(0,\sqrt{2})$ where we have the same second-order periodic solution Eq.(\ref{lpd-br22}). 
\item Along its way, the solution intersects the circle the {\it solid-blue curve} at $(\kappa_1,\kappa_2)=(\sqrt{2-\sqrt{2}},\sqrt{2+\sqrt{2}})$ presented by the {\it purple-solid dot} to the left of $\kappa_1=\kappa_2$ line. Solution Eq.(\ref{lpd-br2}) remain valid as a two breather solution.
\item After the intersection with the circle, the curve reached its maximum where the first derivative of $a_{56}'$ becomes zero at the point $\kappa_1,\kappa_2$=$(\sqrt{2},2)$ identified by the ({\it purple polygon}). At this point with $\kappa_1=\sqrt{2}$ the two-breather solution Eq.(\ref{lpd-br2}) transformed to a first-order periodic solution Eq.(\ref{periodic}).
\item As solution $a_{56}'$ goes further following the line meets with the degenerate line $\kappa_1=\kappa_2$ (solid blue straight line) at $(\kappa_1,\kappa_2)=(\sqrt{2+\sqrt{2}},\sqrt{2+\sqrt{2}})$ where the solution Eq.(\ref{lpd-br22}) becomes invalid. This point is represented by the {\it blue-triangle} near the tip of the line passing through the origin.
\item The end point of $a_{56}'$ is at $(2,\sqrt{2})$ where it meets with the final point of $a_{34}'$ identified by the {\it purple polygon}. At this point the solution Eq.(\ref{lpd-br2}) remain valid and with $\kappa_1=2$ and $\kappa_2=a_{56}'=\sqrt{2}$, Eq.(\ref{lpd-br2}) becomes the first-order periodic solution Eq.(\ref{periodic}).
\end{itemize}

While for the basic NLSE, the degenerate line $\kappa_1=\kappa_2$ intersects the circle at one point as indicated in Fig.(\ref{Fig-Circle}), now for the LPDE, the line intersects three points which also include the ellipses at {\it blue-triangle}, {\it blue star} and the {\it red-circle} in Fig.(\ref{Fig-negative}).

In the recent work \cite{chowdury2017breather}, we gave a detailed analysis of the emergence of solitonic and periodic structures in the solution of LPDE. However, in this present work, we take the opportunity to locate these structures in the parameter space. Comparing the parameter space of basic NLSE and LPDE two-breather solutions, we are now in a position to make several important remarks. Firstly, the region of parameter space for basic NLSE is a circle. In  that circle, a breather never transforms into a periodic or solitonic structure. These observations are supported by several recent works, where it was concluded that  real physical systems described by NLSE  in fact do not allow a breather to become a solitonic or periodic structures \cite{mahnke2012possibility,chowdury2015moving,chowdury2017breatherrule,chowdury2015breathertosoliton}.

Secondly, following the similar observation, it is also concluded that, the emergence of solitonic or periodic structures from breather is only possible when we go beyond the basic NLSE. In our current observations, now in the LPDE system, together with a circle, the parameter space is now defined by intersecting ellipses. These ellipses, indeed, allow the emergence of periodic structures. However, in the limit of infinite period, they become  solitons on a background. 

\begin{figure}[htbp]
\begin{center}
\subfloat[]{\label{Fig17}\includegraphics[width=0.23\textwidth]{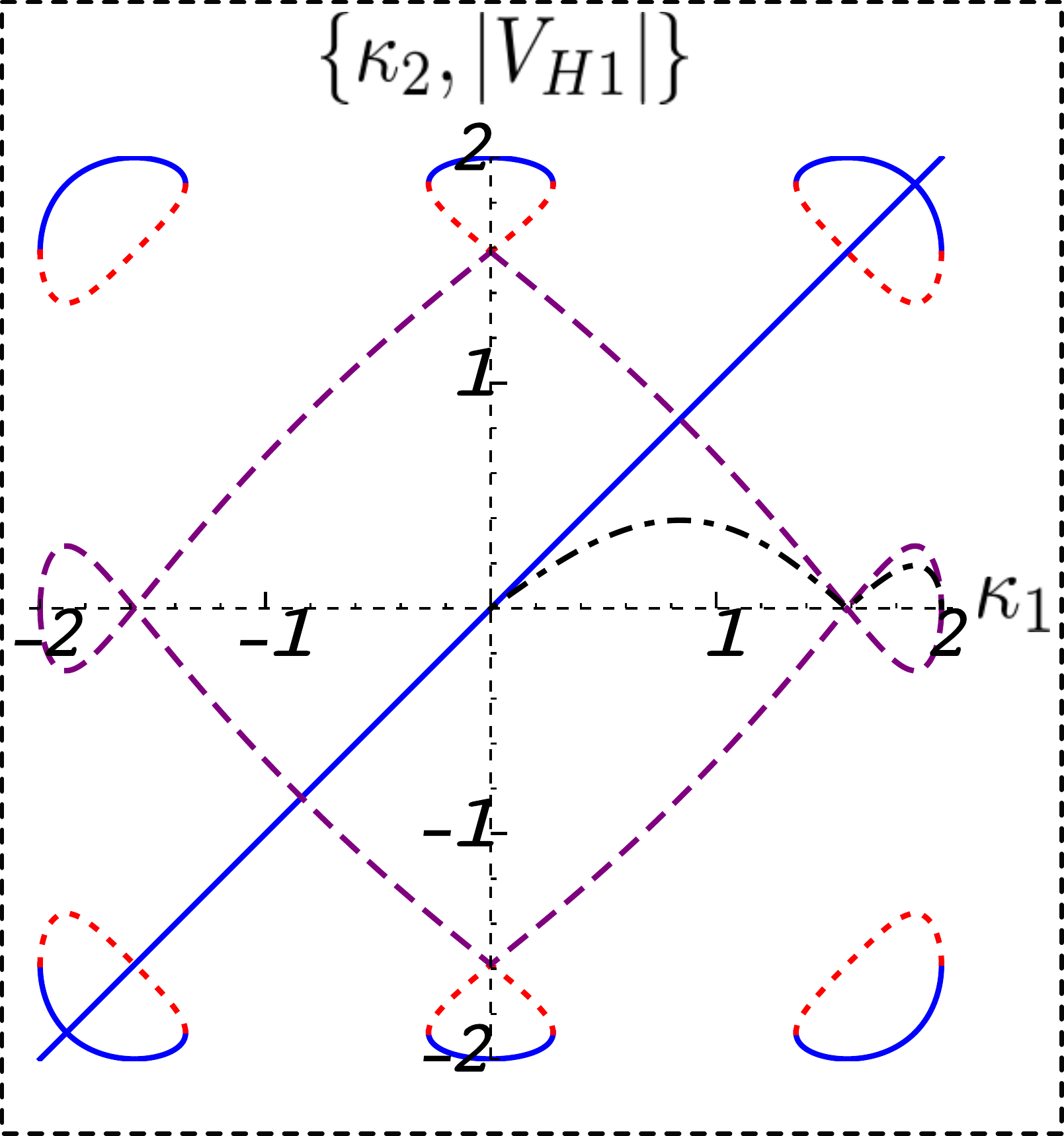}}
\subfloat[]{\label{Fig14}\includegraphics[width=0.23\textwidth]{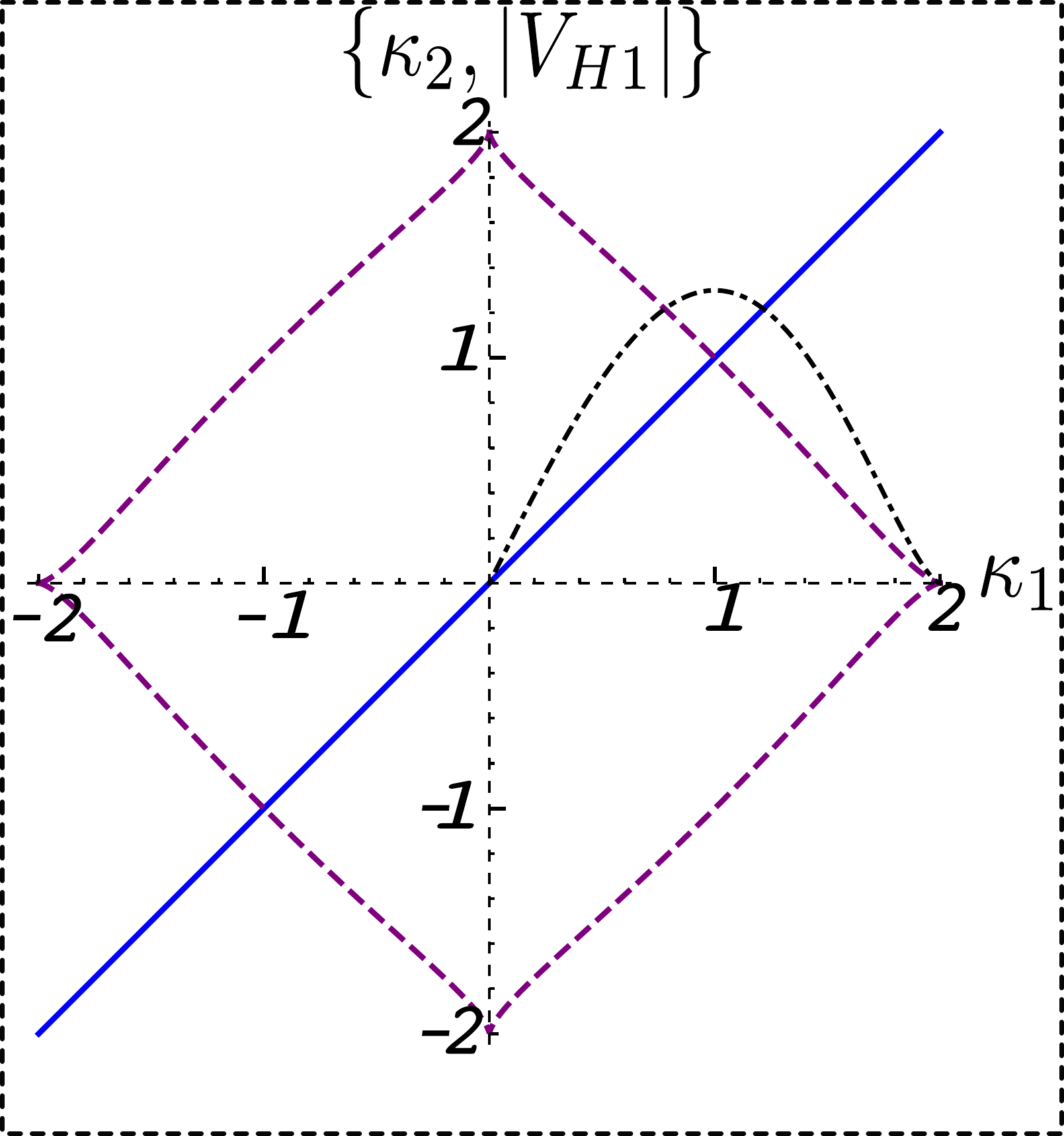}}
\caption{\label{Fig:second-breather}(Color online) Plots of the locus presented in Eq.(\ref{polyp}) with (a) The intersecting circle and ellipses that we get for $\alpha_4=-1/8$ presented in Fig.(\ref{Fig-negative}) now vanish  with $\alpha_4=-1/7$. The parameter space is now moving away to the outward direction. The MI curve, the {\it black dot dashed} line now come in contact at $\kappa_1=\sqrt{3}$ and (b) with $\alpha_4=-1/4$, the locus presented by $a_{12}$ and $a_{34}$ becomes imaginary and we only have two real solutions, $\kappa_2=\kappa_1$, the {\it solid-blue line} and $a_{56}$ presented by the {\it purple-dashed line}. The same MI curve, for $\alpha_4=-1/4$  now touch down at $\kappa_1=2$.}
\end{center}
\end{figure}

From   Eq.(\ref{polyp}), if we deviate from the standard $\alpha_4=-1/8$ value for which we get closed form solutions, we segregate the parameter space. The circle and ellipse no longer exist, as indicated in the Fig.(\ref{Fig17}). For example, we approach a two directional deviation from the value $\alpha_4=-1/8$. 
\begin{itemize}
\item First is from $\alpha_4= -1/8 \to -1/4$ 
\item Second direction is towards $\alpha_4= -1/8 \to -1/12$. 
\end{itemize}

\subsection{case: $\alpha_4= -1/8 \to -1/4$}
\label{right}

Following the deviation towards $\alpha_4= -1/8 \to -1/4$ with $\alpha_4=-1/7$, the blue-half circle in Fig.(\ref{Fig-negative}), now breaks apart and creates an open loop at $y$ axes and another one at the tip of $\kappa_1=\kappa_2$ line as indicated in Fig.(\ref{Fig17}). However, these open loops become closed with the {\it red-dotted line}, the solution $a_{34}$ and makes  a closed region of space. Interestingly, the breather solution only exists on these lines and the region within. In the vicinity of these two separated loops, solutions do not exist, except any point fall on any of the lines presented by these solutions. 

The solution $a_{56}$, the {\it purple-dashed line} now crosses the degenerate line with a reverse slope and hit the $x$ and $y$ axis at $(\kappa_1,\kappa_2)=(\sqrt{3},0)$ and $(0,\sqrt{3})$ respectively. The MI band also come in contact at this point. This is a symmetric process. To see the change of breather solution at these points, we cannot directly set $\kappa_1=0$ which will make the solution indeterminate. With the limit $\kappa_1 \to 0$ at $y$ axis, the breather solution Eq.(\ref{lpd-br2}) becomes the second-order periodic solution Eq.(\ref{lpd-br22}). With $\kappa_1=\sqrt{3}$ at $x$ axis again we get a second-order periodic solution.

However, the line $\kappa_1=\kappa_2$ still intersects solution $a_{12}$, $a_{34}$ and $a_{56}$ at three different points where solution becomes indeterminate.

If we further reduce the value of $\alpha_4$ and move towards the limiting MI value of $\alpha_4=-1/4$, these closed loops and the area within then becomes smaller and at the value of $\alpha_4=-1/4$, we lose solution $a_{12}$ and $a_{34}$ completely (they become imaginary) along with the closed loops, as indicated in the Fig.(\ref{Fig14}). There remains only the solution $a_{56}$, the {\it purple-dashed line}.  Now the MI band come in contact with the $x$ and $y$ axis at $(2,0)$ , $(0,2)$ respectively. Setting $a_{56}$ in Eq.(\ref{lpd-br2}) and if we take $\kappa \to 2$, we get a first-order LPDE rogue ave solution Eq.(\ref{lpd-r-1}). However, the limit $\kappa \to 0$ do not converge here. The solution intersects the degenerate line $\kappa_1=\kappa_2$ at $(1,1)$ where solution becomes indeterminate.

Though the solution $a_{12}$ and $a_{34}$ is imaginary and do not appear on the complex field, yet we have the opportunity to find the condition of the breather on these solution. Interestingly, we revealed that, at $a_{12}$, in the limit of $\kappa_1 \to 0$ and $2$, the Eq.(\ref{lpd-br2}) becomes a LPDE first-order rogue wave solution and a plane wave respectively.

On the other-hand for solution $a_{34}$, while for $\kappa_1 \to 2$ is a plane wave and $\kappa_1 \to 2$ do not converge.

\subsection{case: $\alpha_4= -1/8 \to -1/12$}
\label{left}

\begin{figure}[htbp]
\begin{center}
\subfloat[]{\label{Fig110}\includegraphics[width=0.23\textwidth]{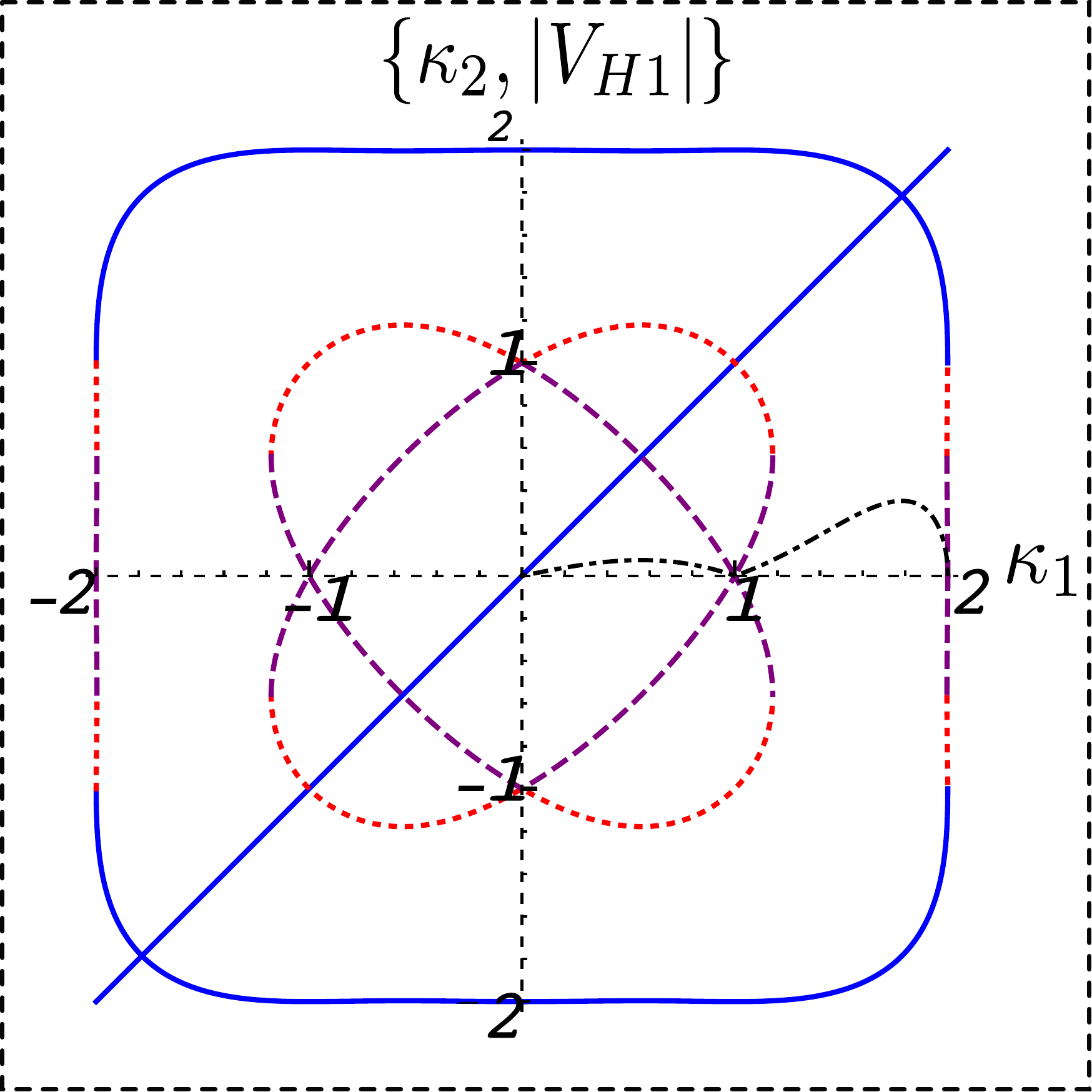}}
\subfloat[]{\label{Fig112}\includegraphics[width=0.23\textwidth]{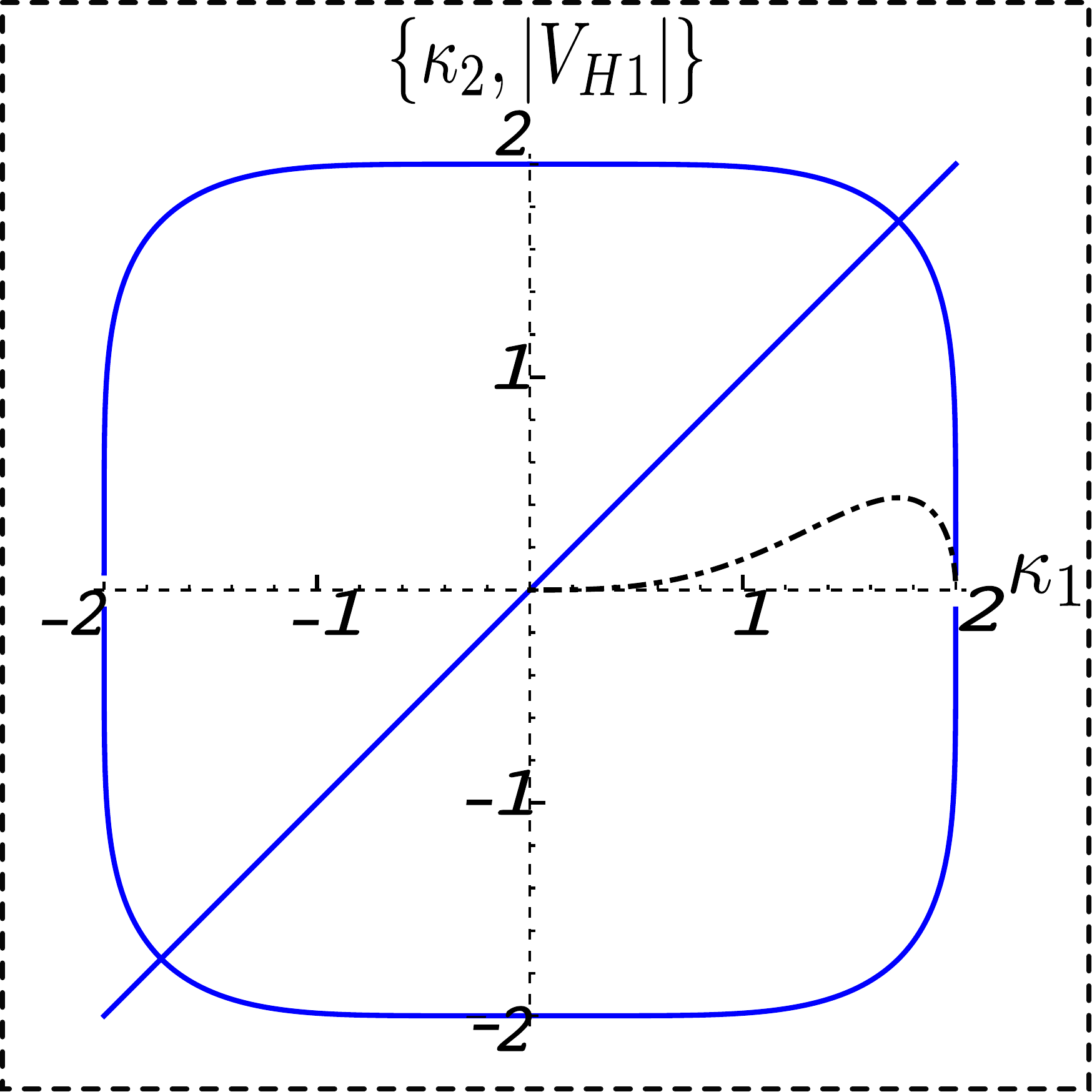}}
\caption{\label{Fig:second-breather}(Color online) Plots of the locus presented in Eq.(\ref{polyp}) with (a) $\alpha_4=-1/10$. Now the parameter space bounded by the locus   moves  inwards. The ellipses representing $a_{34}$ and $a_{56}$ are now shrunk  in area, while the {\it blue circle} representing   $a_{12}$ is reshaping and becoming a square shape. The MI curve presented by {\it black dot dashed line} for $\alpha_4=-1/10$ intersects the $\kappa_1$ line at $1$. Finally, for (b) $\alpha_4=-1/12$, we have two real solutions from Eq.(\ref{polyp}); one is $\kappa_2=\kappa_1$ and the other one is $a_{12}$. The intersecting ellipses have now vanished, making the blue circle to a square. The MI curve now does not touch the $\kappa_1$ line.}
\end{center}
\end{figure}

If we change the value as $\alpha_4= -1/8 \to -1/12$, we will have completely different scenarios, as indicated in the Fig.(\ref{Fig110},\ref{Fig112}). In Fig.(\ref{Fig110}), graphically,   the two ellipses shrink inward while the blue circle   becomes square in size. The solutions $a_{12}$, $a_{34}$ and $a_{56}$ redistribute  themselves and now they intersects each other at fewer points  compared with  Fig.(\ref{Fig-negative}) with $\alpha_4=-1/8$. For example, with $\alpha_4= -1/10$, the solution $a_{12}$, the {\it blue-circle}, now intersects the degenerate line (blue solid line passing through the origin) at $(\kappa_1,\kappa_2)=(\frac{\sqrt{7+\sqrt{33}}}{2},\frac{\sqrt{7+\sqrt{33}}}{2})$ and afterwards meets   the  $a_{34}$ part (small part of a red-dotted line) of the solution at $(\kappa_1,\kappa_2)=(2,1)$. A small parts from both solution $a_{34}$ and $a_{56}$ are positioned vertically at $\kappa_1=2$. Any points that exactly corresponds to $\kappa_1=2$ from these line will give us a first-order periodic solution. For example, solution $a_{12}$ and $a_{34}$ meets the $\kappa_1=2$ at $(2,1)$ therefore give us a first-order periodic solution:
\begin{equation}
\label{periodic5}
\psi=e^{\frac{2 i x}{5}} \left(-1+\frac{1}{2-\sqrt{3} \cos \left(t_{\text{s1}}\right)}\right)
\end{equation}
Any other points will give us a breather with equal gain. The ellipse formed in combination with solution $a_{34}$ and $a_{56}$ indicated the respective parts as {\it red-dotted line} and {\it purple-dashed line} come in contact symmetrically with both $x$ and $y$ axis at $(1,0)$ and (0,1) . The MI band also hit at $(1,0)$ for the same value of $\alpha_4$. At these points with the limit $\kappa \to 0$, we get a second order periodic breather solution. The ellipse intersects the degenerate line at $(\kappa_1,\kappa_2)=(1,1)$. Breather solution Eq.(\ref{lpd-br2}) do not exist outside of the area of the ellipses, except on the lines outside.

Now if we set $\alpha_4=-1/12$ in Eq.(\ref{polyp}), we only have real value for $a_{12}$ the {\it solid-blue curve} and the line $\kappa_1=\kappa_2$. The solution $a_{34}$ and $a_{56}$ becomes imaginary, thus the ellipse vanishes as indicated in Fig.(\ref{Fig112}). With the solution $\kappa_2=a_{12}$ and taking the limit $\kappa_1 \to 0$ we get:
\begin{equation}
\label{soliton}
\psi_1= \left(\frac{4}{1+4  t_{s1}^2} -1\right)e^{i x/2}
\end{equation}
while $\kappa_1 \to 2$ do not converge here. Remarkably, though the solution $a_{34}$ and $a_{56}$ becomes imaginary for $\alpha_4=-1/12$, still we can reveal information of the fate of the two-breather solution Eq.(\ref{lpd-br2}) using $a_{34}$ and $a_{56}$. We found that with the limit $\kappa \to 0$ for both $a_{34}$ and $a_{56}$, we get:

\begin{figure}[htp]
\begin{center}
\subfloat[]{\label{Fig-1}\includegraphics[width=0.23\textwidth]{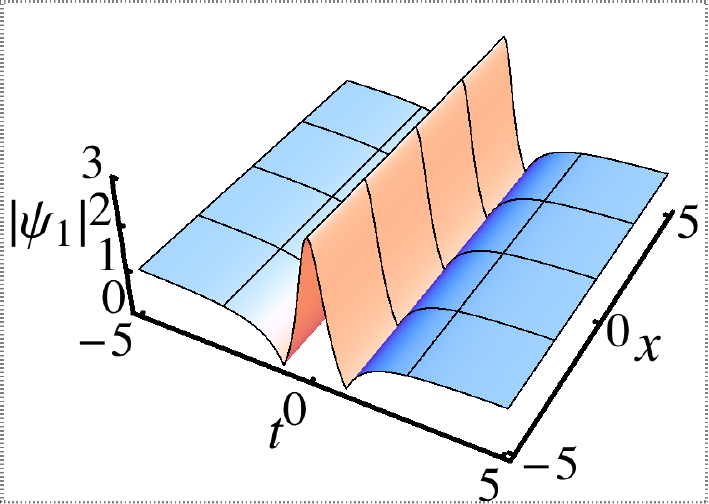}}
\subfloat[]{\label{Fig-2}\includegraphics[width=0.23\textwidth]{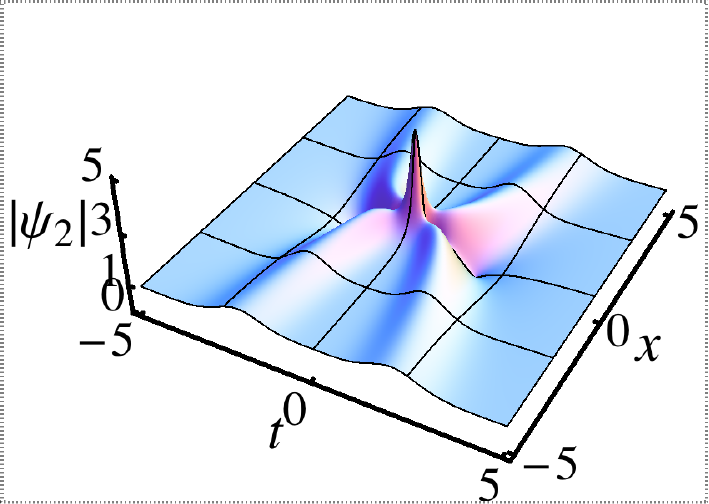}}
\caption{(Color online) Plot of a rational (a) first (b) second order soliton Eq.(\ref{soliton}) and (\ref{soliton2}).
}
\label{unit-circle}
\end{center}
\end{figure}
\begin{eqnarray}
\label{soliton2}
\nonumber G_2&=&-12 \left(-3+24 t^2+16 t^4\right)\\
  H_2&=&-192 \left(1+4 t^2\right) x\\
 \nonumber D_2&=&9+108 t^2+48 t^4+64 t^6+256 x^2 
\end{eqnarray}
with phase factor $e^{ix/2}$ while $\kappa \to 2$ do not converge. At the same time the MI curve do not come in contact with the $\kappa_1$ line at any point. From this point we can say that, the ellipses starts to appear with the MI curve meeting the $\kappa_1$ line and also the breather-soliton features starts to appear at the same time. Now the degenerate line intersects the {\it solid-blue curve} at $(\kappa_1,\kappa_2)=(\sqrt{3},\sqrt{3})$. 

\section{Influence of higher-order terms}

\begin{figure}[htbp]
\begin{center}
\includegraphics[width=0.3\textwidth]{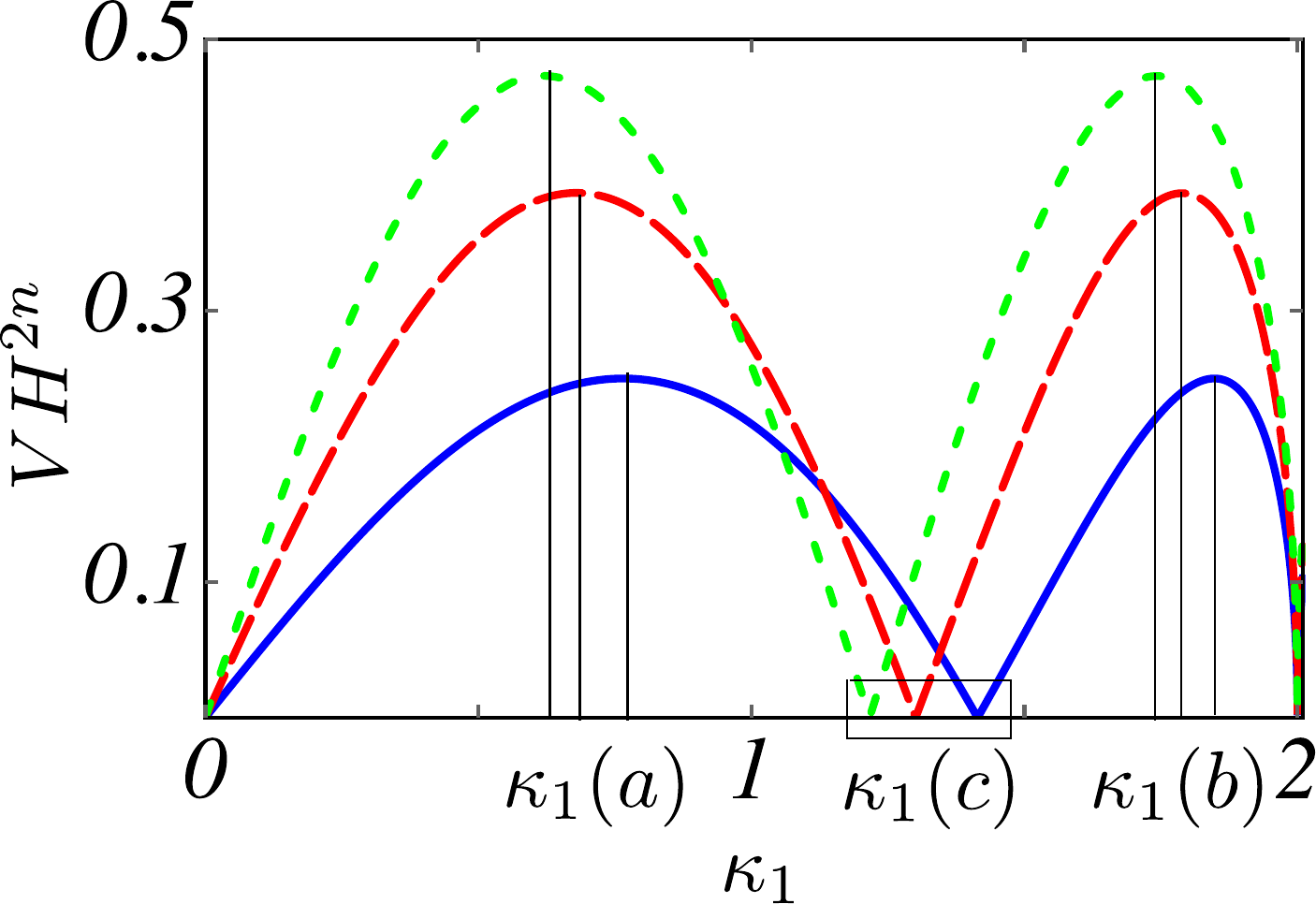}
\caption{\label{Fig:second-breather}(Color online) Plot of MI sub-band for $\alpha^{2n}$ where $n=2,3,4$. $\kappa_1(a)$ and $\kappa_1(b)$ are the corresponding maximum height position. The small box indicates $\kappa_1(c)$ where sub-band hits the $\kappa_1$ line (add NLSE line).  
}
\label{we1}
\end{center}
\end{figure}

Qualitatively, the role of higher-order terms on the equal growth scenarios as well as higher-order MI have similar impact on a general two breather solution. However, for a quantitative investigation we consider to take into account next $6^{th}$  $8^{th}$ and $10^{th}$ order equation. For these case, it is still important to solve Eq.(\ref{lpde}) to get the locuses for equal growth as well as to reveal the region of parameter space for a two breather solution. To solve Eq.(\ref{lpde}), for higher-order cases the growth parameters for a two-breather solution can be given as:
\begin{eqnarray}
\label{(6aa)} V_{Hj}^{(6)}&=&2\delta_j[\alpha_2-\alpha_6(10\kappa_j^2-\kappa_j^4-30)],\\
\label{(8aa)}V_{Hj}^{(8)}&=&2\delta_j[\alpha_2-\alpha_8(70\kappa_j^2-14\kappa_j^4+\kappa_j^6-140)]\\
\label{(10aa)}V_{Hj}^{(10)}&=&2 \delta_j\{\alpha _2+\alpha _{10} [630-420 \kappa _j^2+126 \kappa _j^4\\ \nonumber
&-&18 \kappa _j^6+\kappa _j^8]\}
\end{eqnarray}
where $j=1,2$. The propagation constants now are
\begin{eqnarray*}
\omega^{(6)}&=&2(\alpha_2+10\alpha_6),\\
\omega^{(8)}&=&2(\alpha_2+35\alpha_8),\\
\omega^{(10)}&=&2 \left(\alpha _2+126 \alpha _{10}\right).
\end{eqnarray*}
Even for $4^{th}$ order case, Eq.(\ref{lpde}) is an {\it octic} polynomial and is quite cumbersome to solve analytically (see Appendix. (\ref{appen})).  For a special value of $\alpha_4=-1/8$, the {\it octic} polynomial simplifies greatly as indicated in Eq.(\ref{simplified}) and we get intersecting circles and ellipses. For $4^{th}$ order case, the value of $\alpha_4=-1/8$ is special for number of reason. Firstly, the gain curve hits at $\kappa_1(c)=\sqrt{2}$ at which we get highest gain for NLSE MI band. This clearly indicates that there is a co-relation between the gain curve for NLSE ($2^{nd}$) and LPDE ($4^{th}$). Secondly, at $\sqrt{2}$, we get equal gain for two- sub-bands indicated in Fig.(\ref{we1}). Finally, only this value allows solutions as ellipse and circle.

\begin{table}[http]
\begin{center}
\begin{tabular}{ |p{1.5cm}|p{1.0cm}|p{1.0cm}|p{1.0cm}| p{1.0cm}| p{1.5cm}|}
 \hline
 Equation order &$-1/\alpha_{2n}$ &$ \kappa_1(a)$ &$\kappa_1(b)$&$\kappa_1(c)$&Gain height\\
 \hline
 $4^{th}$& 8 &0.765&1.85&$\sqrt{2}$&0.25\\
 $6^{th}$& 31.90 & 0.679   &1.79&1.30&0.386\\
 $8^{th}$ &127.70 & 0.619&1.74&1.216&0.473\\
 $10^{th}$ &512.52 & 0.574&1.70&1.150&0.533\\
 \hline
\end{tabular}
\end{center}
\caption{List of values of gain band for $\alpha_{2n}$.$\kappa_1(a)$ and $\kappa_1(b)$ are the points corresponding to the maximum gain height where as $\kappa_1(c)$ indicates the points where the sub-bands are hitting at $\kappa_1$ line.}
\label{table1}
\end{table}

\begin{figure}[htbp]
\begin{center}
\includegraphics[width=0.3\textwidth]{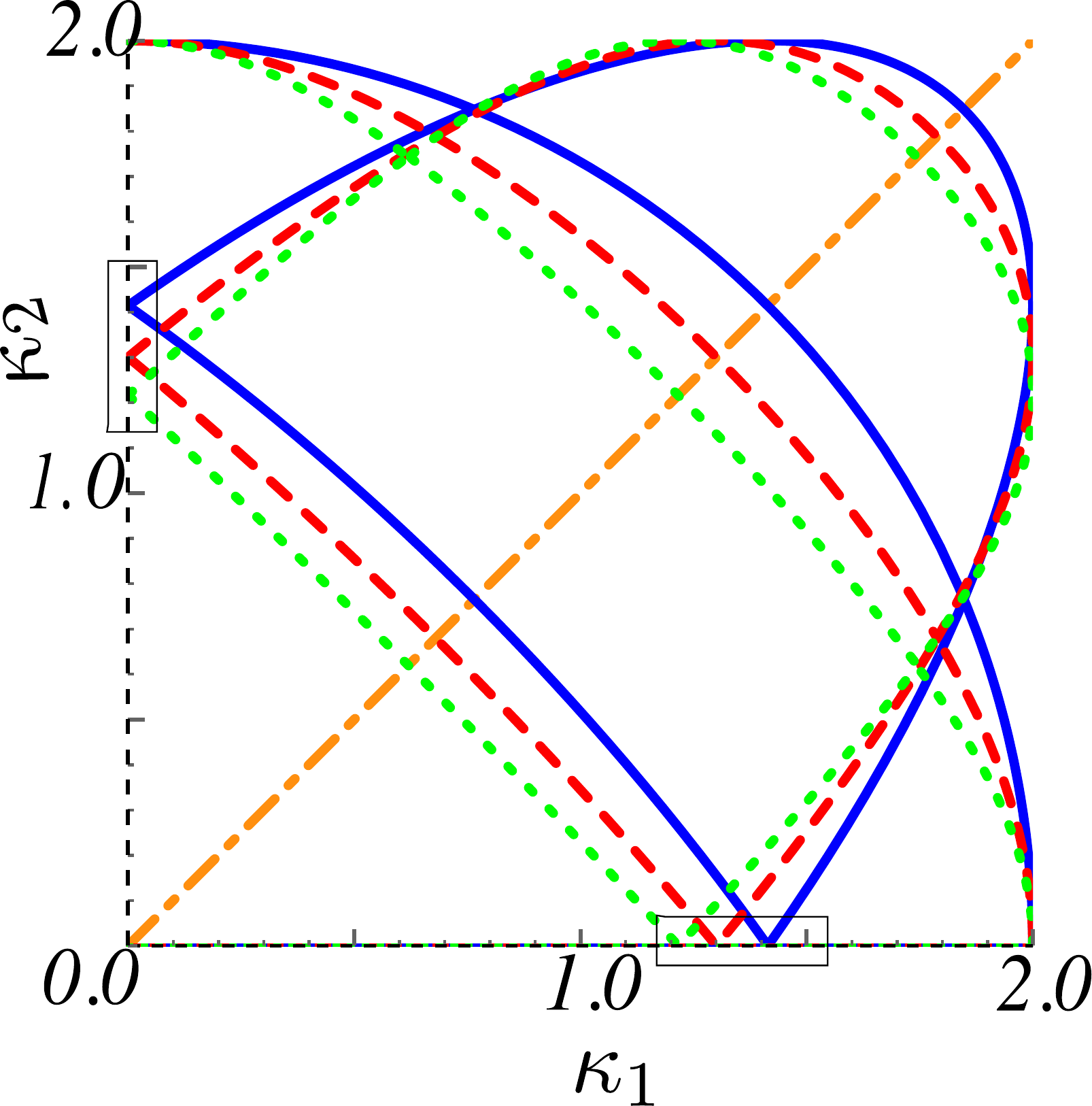}
\caption{\label{Fig:second-breather}(Color online)   Locus for equal growth for a two breather solution for higher-order ($6^{th}$,$8^{th}$ and $10^{th}$) equation. The small boxes at $x$ and $y$ axes corresponds to the same set of points $\kappa_1(c)$ from the table and Fig.(\ref{we1}) .}
\label{we2}
\end{center}
\end{figure}

However, similar values for higher-order equations are not straight forward as they involve polynomials with very high degree. Therefore, we solve them numerically. Indeed, similar values still exist that provides similar locuses as for LPDE, indicated in Fig.(\ref{we2}). We reveal that, though not exactly, but approximately there is a pattern in these values. Such as $\alpha_4=-1/8$, $\alpha_6=-1/(31.90) \approx -1/32$,  $\alpha_8=-1/(127.70) \approx -1/128$ and  $\alpha_{10}=-1/(512.52) \approx -1/512$ $\cdots  \alpha_{2n} \approx -1/ 2^{2n-1}$ for $n=2,3,4....n-1$. Various values obtained from the numerical solutions are presented in the Table (\ref{table1}).

It is clear from the Table (\ref{table1}) that the first maximums points $\kappa_1(a)$ also indicated in Fig.(\ref{we2}) are always $>1$ while the points at second maximum $\kappa_1(b)$ are always $>1$ for any order of equation. These maximum values corresponds to the same $\kappa_1$ values at the intersection points between the degenerate line ({\it orange dot-dashed line}) and the circles and ellipses (show in the figure?). It is clear from the Fig.(\ref{we2}) that with the increasing order of equation, the solution deviates from the region defined by the perfect circle and ellipses as for $\alpha_4$. Now they are transforming to more oblate shapes and moving inward direction.    

However, in the previous sections (\ref{right},\ref{left}), we found that for $-\frac{1}{4 }<\alpha_6<-\frac{1}{12 }$, the parameter space vanishes either inward or outward direction. Similar limit also exist for higher-order equations. These limits are $-\frac{1}{12 }<\alpha_6<  -\frac{1}{60 }$ and $-\frac{1}{40 }<\alpha_8<  -\frac{1}{280}$ for $6^{th}$ and $8^{th}$ order equation respectively.

\section{Conclusions}
Our main results are following:
\begin{enumerate}

\item The two-breather solution of  the NLSE is studied rigorously and a large volume of literature is available on this topic. It has many interesting properties along with its various forms. In most cases, its multiple forms of solution arise from the precise manipulation of the modulation frequencies $\kappa_1$ and $\kappa_2$. In our analysis, we showed that the choice of these parameter values are bounded by a circle with radius $2$. 

\item In a two breather solution, each breather has its own growth rate such as $\delta_1$ and $\delta_2$. One of the interesting form of the two-breather solution of NLSE is the `degenerate solution ' where parameter $\kappa_1=\kappa_2$. Usually, following a special technique, the solution is transformed to an expression with a single modulation frequency, though it is still a two breather solution with a common growth rate. Rather than defining the condition $\kappa_1=\kappa_2$, we can   obtain it as one of the solutions of equal growth rates, $\delta_1=\delta_2$. The solutions of the equal growth condition mark  the loci where the growth of a two breather solution will be equal. 

We revealed that for a NLSE two-breather, we will have equal growth on a line passing through the origin and on the boundary of a circle of radius $2$. We also showed that, in the zero frequency limit on these loci, in one case we get a second order rational solution and in other case a first order one. 

\item We showed that if we go beyond the basic NLSE, for the positive value of the coefficient $\alpha_4$, the choice of parameter $\kappa_1$ and $\kappa_2$ is bounded by a circle and growth is equal at the boundary of the circle and for $\alpha_4 \simeq 1$, the parameter space reshaped to a square with high growth rate and equal growth occurs on the sides of the square.

\item However, for negative values of  $\alpha_4$ within a specific range, the entire parameter space is bounded by an intersecting circle and two ellipses. We explicitly derived and showed that, for a LPDE two-breather solution, the growth rate is equal at the boundary of the circle and ellipses. In addition, we showed that, with the variation of the negative value of $\alpha_4$, the parameter space also varies following a set of specific rules.
\end{enumerate}

\section*{Acknowledgements}

\section*{Appendix: Solutions of Eq.(\ref{lpde}) with arbitrary $\alpha_4$:}
\label{appen}
Keeping $\alpha_2=1/2$ with  $\alpha_4 \ne 0$ the solutions are: 
\begin{eqnarray}
 \label{polyp}
\nonumber a_{12}&=&\pm \frac{1}{\sqrt{6}} \sqrt{\frac{m_1^{2/3}+m_2+2 \sqrt[3]{m_1} m_3}{\sqrt[3]{m_1} \alpha _4^2}}
\\
\nonumber a_{34}&=&\pm \frac{1}{2 \sqrt{3}} \sqrt\{\frac{1}{\sqrt[3]{m_1} \alpha _4^2} [(-1-i \sqrt{3}) m_1^{\frac{2}{3}}\\
 &-&(1-i \sqrt{3}) m_2+4 m_1^{\frac{1}{3}} m_3]\}
 \\
\nonumber a_{56}&=&\pm \frac{1}{2 \sqrt{3}} \sqrt \{\frac{1}{\sqrt[3]{m_1} \alpha _4^2} [(-1+i \sqrt{3}) m_1^{\frac{2}{3}}\\
\nonumber &-&(1+i \sqrt{3}) m_2+4 m_1^{\frac{1}{3}} m_3]\}
\end{eqnarray}
where $m_1, m_2$ and $m_3$ can be given as 
\begin{eqnarray*}
m_1&=&\alpha _4^3 \{-1+2 \alpha _4 [-3 \left(2+\kappa _1^2\right)\\
&+&6 \alpha _4 \left(-4-28 \kappa _1^2+5 \kappa _1^4\right)\\
&-&8 \alpha _4^2 \left(4+186 \kappa _1^2-60 \kappa _1^4+5 \kappa _1^6\right)]\}\\
&+&6 \sqrt{3} \sqrt\{\alpha _4^8 \kappa _1^2 \left(-4+\kappa _1^2\right) \\
&\times&[-1+4 \alpha _4 (-6-\kappa _1^2+\alpha _4 \left(-48-68 \kappa _1^2+11 \kappa _1^4\right)\\
&-&8 \alpha _4^2 \left(20+134 \kappa _1^2-40 \kappa _1^4+3 \kappa _1^6\right)\\
&+&16 \alpha _4^3 (-12-312 \kappa _1^2+136 \kappa _1^4-20 \kappa _1^6+\kappa _1^8))]\}
\\
m_2&=&\alpha _4^2+4 \alpha _4^3 \left(2+\kappa _1^2\right)-8 \alpha _4^4 \left(-2-8 \kappa _1^2+\kappa _1^4\right)
\\
m_3&=&\alpha _4 \left(1-\alpha _4 \left(-16+\kappa _1^2\right)\right)
\end{eqnarray*}

\bibliographystyle{unsrt}
\bibliography{amdad-updated-2}

\begin{thebibliography}{10}

\bibitem{benjamin1967disintegration}
T.~B. Benjamin and J.~E. Feir.
\newblock The disintegration of wave trains on deep water part 1. theory.
\newblock {\em J. Fluid Mechanics}, {\bf 27}(03):417--430, (1967).

\bibitem{bespalov1966filamentary}
V.~I. Bespalov and V.~I. Talanov.
\newblock Filamentary structure of light beams in nonlinear liquids.
\newblock {\em ZhETF Pisma Redaktsiiu}, {\bf 3}:471, (1966).

\bibitem{van2001experimental}
G.~Van~Simaeys, Ph. Emplit, and M.~Haelterman.
\newblock {Experimental demonstration of the Fermi-Pasta-Ulam recurrence in a
  modulationally unstable optical wave}.
\newblock {\em Phys. Rev. Lett.}, {\bf 87}(3):033902, (2001).

\bibitem{tai1986observation}
K.~Tai, A.~Hasegawa, and A.~Tomita.
\newblock Observation of modulational instability in optical fibers.
\newblock {\em Phys. Rev. Lett.}, {\bf 56}(2):135, 1986.

\bibitem{greer1989generation}
E.~J. Greer, D.~M. Patrick, P.~G.~J. Wigley, and J.~R. Taylor.
\newblock {Generation of 2 THz repetition rate pulse trains through induced
  modulational instability}.
\newblock {\em Electronics Letters}, {\bf 25}(18):1246--1248, (1989).

\bibitem{agrawal1987modulation}
G.~P. Agrawal.
\newblock Modulation instability induced by cross-phase modulation.
\newblock {\em Phys. Rev. Lett.}, {\bf 59}(8):880, (1987).

\bibitem{demircan2005supercontinuum}
A.~Demircan and U.~Bandelow.
\newblock Supercontinuum generation by the modulation instability.
\newblock {\em Opt. Commun.}, {\bf 244}(1):181--185, (2005).

\bibitem{dudley2006supercontinuum}
John~M. Dudley, Go{\"e}ry Genty, and St{\'e}phane Coen.
\newblock Supercontinuum generation in photonic crystal fiber.
\newblock {\em Rev. Modern Phys.}, {\bf 78}(4):1135, (2006).

\bibitem{travers2008visible}
J.~C. Travers, A.~B. Rulkov, B.~A. Cumberland, S.~V. Popov, and J.~R. Taylor.
\newblock {Visible supercontinuum generation in photonic crystal fibers with a
  400W continuous wave fiber laser}.
\newblock {\em Optics Express}, {\bf 16}(19):14435--14447, (2008).

\bibitem{dudley2009modulation}
J.~M. Dudley, G.~Genty, F.~Dias, B.~Kibler, and N.~Akhmediev.
\newblock {Modulation instability, Akhmediev Breathers and continuous wave
  supercontinuum generation}.
\newblock {\em Optics Express}, {\bf 17}(24):21497--21508, (2009).

\bibitem{genty2010akhmediev}
G.~Genty, F.~Dias, B.~Kibler, N.~Akhmediev, and J.~M. Dudley.
\newblock Akhmediev breather dynamics and the nonlinear modulation instability
  spectrum.
\newblock In {\em SPIE Photonics Europe}, pages 772812--772812. International
  Society for Optics and Photonics, (2010).

\bibitem{Solli:07:Nature}
D.~R. Solli, C.~Ropers, P.~Koonath, and B.~Jalali.
\newblock {Optical rogue waves}.
\newblock {\em Nature}, {\bf 450}:1054, (2007).

\bibitem{akhmediev1986modulation}
N.~Akhmediev and V.~I. Korneev.
\newblock Modulation instability and periodic solutions of the nonlinear
  {S}chr{\"o}dinger equation.
\newblock {\em Theor. Math. Phys.}, {\bf 69}(2):1089--1093, (1986).

\bibitem{trillo1991dynamics}
S.~Trillo and Stefan Wabnitz.
\newblock Dynamics of the nonlinear modulational instability in optical fibers.
\newblock {\em Optics Letters}, {\bf 16}(13):986--988, (1991).

\bibitem{trillo1994nonlinear}
S.~Trillo, S.~Wabnitz, and T.A.B. Kennedy.
\newblock Nonlinear dynamics of dual-frequency-pumped multiwave mixing in
  optical fibers.
\newblock {\em Physical Review A}, {\bf 50}(2):1732, (1994).

\bibitem{Akhmediev:09:PLA2}
N.~Akhmediev, J.~M. Soto-Crespo, and A.~Ankiewicz.
\newblock {Extreme waves that appear from nowhere: On the nature of rogue
  waves}.
\newblock {\em Physics Letters A}, {\bf 373}(25):2137--2145, (2009).

\bibitem{Tajiri:98:PRE}
M.~Tajiri and Y.~Watanabe.
\newblock {Breather solutions to the focusing nonlinear Schr\"odinger
  equation}.
\newblock {\em Physical Review E}, {\bf 57}:3510 -- 3519, (1998).

\bibitem{Park:99:PRL}
Q-Han Park and H.~J. Shin.
\newblock {Parametric control of soliton light traffic by cw traffic light}.
\newblock {\em Physical Review Letters}, {\bf 82}:4432 -- 4435, (1999).

\bibitem{li2004modulation}
Lu~Li, Zhonghao Li, Shuqing Li, and Guosheng Zhou.
\newblock Modulation instability and solitons on a cw background in
  inhomogeneous optical fiber media.
\newblock {\em Optics Communications}, {\bf 234}(1):169--176, (2004).

\bibitem{Zakharov:13:PRL}
V.~E. Zakharov and A.~A. Gelash.
\newblock {Nonlinear Stage of Modulation Instability}.
\newblock {\em Physical Review Letters}, {\bf 111}:054101, (2013).

\bibitem{akhmediev1985generation}
N.~Akhmediev, V.~Eleonsky, and N.~Kulagin.
\newblock Generation of periodic trains of picosecond pulses in an optical
  fiber: exact solutions.
\newblock {\em Sov. Phys. JETP}, {\bf 62}(5):894--899, (1985).

\bibitem{akhmediev1988n}
N.N. Akhmediev, V.I. Korneev, and N.V. Mitskevich.
\newblock N-modulation signals in a single-mode optical waveguide under
  nonlinear conditions.
\newblock {\em Journal of Experimental and Theoretical Physics}, {\bf
  67}(1):89, (1988).

\bibitem{backus19970}
Sterling Backus, Charles~G. Durfee~III, Gerard Mourou, Henry~C. Kapteyn, and
  Margaret~M. Murnane.
\newblock 0.2-tw laser system at 1khz.
\newblock {\em Optics Letters}, {\bf 22}(16):1256--1258, (1997).

\bibitem{tao2012multisolitons}
Yongsheng Tao and Jingsong He.
\newblock Multisolitons, breathers, and rogue waves for the {H}irota equation
  generated by the {D}arboux transformation.
\newblock {\em Physical Review E}, {\bf 85}(2):026601, (2012).

\bibitem{akhmediev1997solitons}
N.~Akhmediev and A.~Ankiewicz.
\newblock {\em Solitons: nonlinear pulses and beams}.
\newblock Chapman \& Hall, London, 1997.

\bibitem{kibler2012observation}
Bertrand Kibler, Julien Fatome, Christophe Finot, Guy Millot, Go{\"e}ry Genty,
  Benjamin Wetzel, N.~Akhmediev, Fr{\'e}d{\'e}ric Dias, and John~M. Dudley.
\newblock Observation of {K}uznetsov-{M}a soliton dynamics in optical fibre.
\newblock {\em Scientific Reports}, {\bf 2}, (2012).

\bibitem{dudley2014instabilities}
John~M. Dudley, Fr{\'e}d{\'e}ric Dias, Miro Erkintalo, and Go{\"e}ry Genty.
\newblock Instabilities, breathers and rogue waves in optics.
\newblock {\em Nature Photonics}, {\bf 8}(10):755--764, (2014).

\bibitem{chabchoub2014hydrodynamics}
A.~Chabchoub, B.~Kibler, J.~M. Dudley, and N.~Akhmediev.
\newblock Hydrodynamics of periodic breathers.
\newblock {\em Philosophical Transactions of the Royal Society A: Mathematical,
  Physical and Engineering Sciences}, {\bf 372}(2027):20140005, (2014).

\bibitem{Dysthe:99:PS}
K.~B. Dysthe and K.~Trulsen.
\newblock {Note on Breather type Solutions of the NLS as models for
  freak-waves}.
\newblock {\em Physica Scripta}, {\bf T82}:48--52, (1999).

\bibitem{Peregrine:83:JAMS}
D.H. Peregrine.
\newblock {Water waves, nonlinear Schr{\"o}dinger equations and their
  solutions}.
\newblock {\em Journal of the Australian Mathematical Society Series B}, {\bf
  25}(01):16--43, (1983).

\bibitem{henderson1999unsteady}
K.L. Henderson, D.H. Peregrine, and J.W. Dold.
\newblock Unsteady water wave modulations: fully nonlinear solutions and
  comparison with the nonlinear {S}chr{\"o}dinger equation.
\newblock {\em Wave motion}, {\bf 29}(4):341--361, (1999).

\bibitem{Kharif:09:Book}
C.~Kharif, E.~Pelinovsky, and A.~Slunyaev.
\newblock {\em {Rogue waves in the ocean}}.
\newblock Springer, (2009).

\bibitem{Shrira:10:JEM}
V.I. Shrira and V.V. Georjaev.
\newblock {What makes the Peregrine soliton so special as a prototype of freak
  waves?}
\newblock {\em Journal of Engineering Mathematics}, {\bf 67}(1):11--22, (2010).

\bibitem{baronio2015baseband}
Fabio Baronio, Shihua Chen, Philippe Grelu, Stefan Wabnitz, and Matteo
  Conforti.
\newblock {Baseband modulation instability as the origin of rogue waves}.
\newblock {\em Physical Review A}, {\bf 91}(3):033804, (2015).

\bibitem{bandelow2012persistence}
Uwe Bandelow and Nail Akhmediev.
\newblock {Persistence of rogue waves in extended nonlinear Schr{\"o}dinger
  equations: integrable Sasa--Satsuma case}.
\newblock {\em Physics Letters A}, {\bf 376}(18):1558--1561, (2012).

\bibitem{Kedziora:12:PRE1}
D.~J. Kedziora, A.~Ankiewicz, and N.~Akhmediev.
\newblock {Second-order nonlinear Schr\"odinger equation breather solutions in
  the degenerate and rogue wave limits}.
\newblock {\em Physical Review E}, {\bf 85}:066601, Jun (2012).

\bibitem{Akhmediev:09:PRE}
N.~Akhmediev, A.~Ankiewicz, and J.~M. Soto-Crespo.
\newblock {Rogue waves and rational solutions of the nonlinear Schr\"odinger
  equation}.
\newblock {\em Physical Review E}, {\bf 80}(2):026601, (2009).

\bibitem{Erkintalo:11:PRL}
M.~Erkintalo, K.~Hammani, B.~Kibler, C.~Finot, N.~Akhmediev, J.~M. Dudley, and
  G.~Genty.
\newblock {Higher-Order Modulation Instability in Nonlinear Fiber Optics}.
\newblock {\em Physical Review Letters}, {\bf 107}:253901, (2011).

\bibitem{hammani2011peregrine}
Kamal Hammani, Bertrand Kibler, Christophe Finot, Philippe Morin, Julien
  Fatome, John~M Dudley, and Guy Millot.
\newblock Peregrine soliton generation and breakup in standard
  telecommunications fiber.
\newblock {\em Optics letters}, 36(2):112--114, 2011.

\bibitem{erkintalo2012seeded}
Miro Erkintalo, Kamal Hammani, Bertrand Kibler, Christophe Finot, Nail
  Akhmediev, John~M Dudley, and Go{\"e}ry Genty.
\newblock Seeded and spontaneous higher-order modulation instability.
\newblock In {\em Frontiers in Optics}, pages FW3A--45. Optical Society of
  America, 2012.

\bibitem{kimmoun2017nonconservative}
O~Kimmoun, HC~Hsu, B~Kibler, and A~Chabchoub.
\newblock Nonconservative higher-order hydrodynamic modulation instability.
\newblock {\em Physical Review E}, 96(2):022219, 2017.

\bibitem{porsezian1992integrability}
K.~Porsezian, M.~Daniel, and M.~Lakshmanan.
\newblock {On the integrability aspects of the one-dimensional classical
  continuum isotropic biquadratic Heisenberg spin chain}.
\newblock {\em J. Math. Phys.}, 33(5):1807--1816, 1992.

\bibitem{wang2013breather}
L.~H. Wang, K.~Porsezian, and J.~S. He.
\newblock {Breather and rogue wave solutions of a generalized nonlinear
  Schr{\"o}dinger equation}.
\newblock {\em Phys. Rev. E}, 87(5):053202, 2013.

\bibitem{agrawal2011nonlinear}
G.~P. Agrawal.
\newblock Nonlinear fiber optics: its history and recent progress [invited].
\newblock {\em JOSA B}, 28(12):A1--A10, 2011.

\bibitem{hook1993ultrashort}
Anders H{\"o}{\"o}k and Magnus Karlsson.
\newblock Ultrashort solitons at the minimum-dispersion wavelength: effects of
  fourth-order dispersion.
\newblock {\em Optics Letters}, {\bf 18}(17):1388--1390, (1993).

\bibitem{chowdury2017breather}
A.~Chowdury, W.~Krolikowski, and N.~Akhmediev.
\newblock {Breather solutions of a fourth-order nonlinear Schr{\"o}dinger
  equation in the degenerate, soliton, and rogue wave limits}.
\newblock {\em Phys. Rev. E}, 96(4):042209, 2017.

\bibitem{Ankiewicz:11:PLA}
A.~Ankiewicz, D.~J. Kedziora, and N.~Akhmediev.
\newblock {Rogue wave triplets}.
\newblock {\em Physics Letters A}, {\bf 375}(28-29):2782 -- 2785, (2011).

\bibitem{Kedziora:11:PRE}
D.~J. Kedziora, A.~Ankiewicz, and N.~Akhmediev.
\newblock {Circular rogue wave clusters}.
\newblock {\em Physical Review E}, {\bf 84}:056611, (2011).

\bibitem{Kedziora:12:PRE2}
D.~J. Kedziora, A.~Ankiewicz, and N.~Akhmediev.
\newblock {Triangular rogue wave cascades}.
\newblock {\em Physical Review E}, {\bf 86}(5):056602, (2012).

\bibitem{Kedziora:13:PRE}
D.~J. Kedziora, A.~Ankiewicz, and N.~Akhmediev.
\newblock {Classifying the hierarchy of nonlinear Schr{\"o}dinger equation
  rogue wave solutions}.
\newblock {\em Physical Review E}, {\bf 88}(1):013207, (2013).

\bibitem{wang2016breather}
Lei Wang, Jian-Hui Zhang, Zi-Qi Wang, Chong Liu, Min Li, Feng-Hua Qi, and Rui
  Guo.
\newblock Breather-to-soliton transitions, nonlinear wave interactions, and
  modulational instability in a higher-order generalized nonlinear
  schr{\"o}dinger equation.
\newblock {\em Physical Review E}, 93(1):012214, 2016.

\bibitem{yang2017breathers}
Jin-Wei Yang, Yi-Tian Gao, Chuan-Qi Su, Qi-Min Wang, and Zhong-Zhou Lan.
\newblock Breathers and rogue waves in a heisenberg ferromagnetic spin chain or
  an alpha helical protein.
\newblock {\em Communications in Nonlinear Science and Numerical Simulation},
  48:340--349, 2017.

\bibitem{chowdury2015breather}
A.~Chowdury, D.~J. Kedziora, A.~Ankiewicz, and N.~Akhmediev.
\newblock {Breather solutions of the integrable quintic nonlinear
  Schr{\"o}dinger equation and their interactions}.
\newblock {\em Physical Review E}, {\bf 91}(2):022919, (2015).

\bibitem{chowdury2015breathertosoliton}
A.~Chowdury, D.~J. Kedziora, A.~Ankiewicz, and N.~Akhmediev.
\newblock {Breather-to-soliton conversions described by the quintic equation of
  the nonlinear Schr{\"o}dinger hierarchy}.
\newblock {\em Phys. Rev. E}, 91(3):032928, 2015.

\bibitem{chowdury2015moving}
A.~Chowdury, A.~Ankiewicz, and N.~Akhmediev.
\newblock Moving breathers and breather-to-soliton conversions for the hirota
  equation.
\newblock {\em Proc. R. Soc. A}, 471(2180):20150130, 2015.

\bibitem{ankiewicz2016superposition}
A~Ankiewicz and A~Chowdury.
\newblock Superposition of solitons with arbitrary parameters for higher-order
  equations.
\newblock {\em Zeitschrift f{\"u}r Naturforschung A}, 71(7):647--656, 2016.

\bibitem{mahnke2012possibility}
C.~Mahnke and F.~Mitschke.
\newblock {Possibility of an Akhmediev breather decaying into solitons}.
\newblock {\em Phys. Rev. A}, {\bf 85}(3):033808, (2012).

\bibitem{chowdury2017breatherrule}
A.~Chowdury and W.~Krolikowski.
\newblock Breather-to-soliton transformation rules in the hierarchy of
  nonlinear schr{\"o}dinger equations.
\newblock {\em Phys. Rev. E}, 95(6):062226, 2017.

\end{thebibliography}
\end{document}